\documentclass[12pt]{iopart}

\usepackage{graphicx}% Include figure files
\usepackage{mathrsfs}
\usepackage{arydshln}
\usepackage{accents}
\usepackage{color}

\usepackage[normalem]{ulem}

\usepackage{iopams}

\newtheorem{remark}{{\bf Remark}}
\newtheorem{example}{{\bf Example}}

\def \ad{\mathrm{ad}}

\newcommand{\qed}{\nobreak \ifvmode \relax \else
      \ifdim\lastskip<1.5em \hskip-\lastskip
      \hskip1.5em plus0em minus0.5em \fi \nobreak
      \vrule height0.75em width0.5em depth0.25em\fi}

\newcommand{\dbracket}[2]{ {[}\mkern-2mu{[} #1 , #2 {]}\mkern-2mu{]} }
\newcommand{\dpoisson}[2]{ {\{}\mkern-5mu{\{} \, #1 , #2 \,{\}}\mkern-5mu{\}} }
\newcommand{\dangle}[2]{ {\langle}\mkern-2mu{\langle} \, #1 , #2 \,{\rangle}\mkern-2mu{\rangle} }

\begin{document}

\title[Nambu mechanics]{Nambu mechanics viewed as a Clebsch parameterized Poisson algebra
\\
--- toward canonicalization and quantization}
% \title{Nambu Mechanics Induced by a Lie-Poisson Algebra}

%%%% To generate auto affiliation numbers please use \author{}\affil{} command

\author{Zensho Yoshida}
\address{National Institute for Fusion Science
\\
Oroshi, Toki City, Gifu 509-5292, Japan
}
% \\ \email{yoshidazensho@gmail.com}}

%%% To include the collaborator name... Please use the command "\collaborator"
%%% For example: \collaborator{ATLAS Collaboration}

\begin{abstract}%
In his pioneering paper [Phys. Rev. E \textbf{7}, 2405 (1973)], Nambu proposed the idea of multiple Hamiltonian systems.
The explicit example examined there is equivalent to the $\mathfrak{so}(3)$ Lie-Poisson system,
which represents noncanonical Hamiltonian dynamics with a Casimir;
the Casimir corresponds to the second Hamiltonian of Nambu's formulation.
The vortex dynamics of ideal fluid, while it is infinite dimensional, has a similar structure, in which the Casimir is the helicity.
These noncanonical Poisson algebras are derived by the reduction,
i.e., restricting the phase space to some submanifold embedded in the canonical phase space.
We may reverse the reduction to canonicalize some Nambu dynamics, i.e., view the Nambu dynamics as the subalgebra of a larger canonical Poisson algebra.
Then, we can invoke the standard corresponding principle for quantizing the canonicalized system.
The inverse of the reduction, i.e., representing the noncanonical variables by some canonical variables may be said ``Clebsch parameterization'' following the fluid mechanical example.
\end{abstract}

% \subjectindex{A00, A60}

\maketitle

%%%%%%%%%%%%%%%%%%%%%%%%%%%%%%%%%%%%%%%%%%%%%%%%%%%%%%%%%%%
\section{Introduction}
\label{sec:introduction}
In this paper, we discuss the Clebsch parameterization of noncanonical Poisson manifolds for the canonicalization and quantization of a special class of Nambu dynamics.
In his original paper\,\cite{Nambu}, 
Nambu introduced the ``generalized Hamiltonian dynamics'' that is dictated by multiple (say $m$) Hamiltonians.
By generalizing the usual Poisson bracket to an appropriate antisymmetric product of $m+1$ observables, such as $\{ G, H_1,\cdots,H_m\}$, one may formulate an \emph{incompressible} (i.e., volume conserving) phase-space velocity 
\begin{equation}
\dot{z}_j = \{z_j, H_1,\cdots,H_m\} ,
\label{Nambu_dynamics}
\end{equation}
satisfying Liouville's theorem ($z_j$ denotes the phase space coordinate).
Then, one obtains a generalized Boltzmann distribution such that 
\begin{equation}
f \propto \rme^{-(\beta_1 H_1 +\cdots+\beta_m H_m)}.
\label{generalized_Boltzmann}
\end{equation}
The model of Euler top was invoked to illustrate how such a generalized Poisson bracket applies to physics:
In a 3-dimensional phase space $M=\mathbb{R}^3$,
we define
\begin{equation}
\{G, H_1, H_2\}= \nabla G \cdot (\nabla H_1)\times(\nabla H_2) .
\label{Nambu_bracket}
\end{equation}
Inserting $H_1 = \frac{1}{2} [ (z_1)^2/I_1 + (z_2)^2/I_2 + (z_3)^2/I_3]$ (kinetic energy of a rigid body with moments of inertia $I_1, I_2$ and $I_3$)
and $H_2= \frac{1}{2} [ (z_1)^2 + (z_2)^2 + (z_3)^2]$ (square of the angular momentum $\bi{z}\in M$),
(\ref{Nambu_dynamics}) reads the equation of motion of the Euler top.
Evidently, we find $\{H_1, H_1, H_2\}=\{H_2, H_1, H_2\}=0$, implying the conservation of the magnitude of angular momentum, as well as the kinetic energy.

From the foregoing sketch of the Nambu bracket, we notice the following well-known relations:
\begin{enumerate}
\item[(i)]
The bracket (\ref{Nambu_bracket}), for the specific choice of $H_2= \frac{1}{2} |\bi{z}|^2$, can be viewed as a Lie-Poisson bracket such that
\begin{equation}
\{ G , H_1\}_{\mathfrak{so}(3)} = \langle [\partial_{\bi{z}}G, \partial_{\bi{z}}H_1]_{\mathfrak{so}(3)}, \bi{z} \rangle ,
\label{so(3)_Lie-Poisson}
\end{equation}
where $[\bi{x},\bi{y}]_{\mathfrak{so}(3)}=\bi{x}\times\bi{y}$ is the $\mathfrak{so}(3)$ Lie bracket\,\cite{Morrison1998}
(see Sec.\,\ref{sec:Lie-Poisson} for the systematic derivation of a Lie-Poisson bracket from a Lie algebra).
In fact, inserting $H_2=\frac{1}{2}|\bi{z}|^2$ in (\ref{Nambu_bracket}), we observe
\begin{equation}
\{ G, H_1, H_2 \} = \{ G , H_1\}_{\mathfrak{so}(3)}.
\quad (\forall G,\, H_1 ).
\label{so(3)_Lie-Poisson-2}
\end{equation}
Evidently, $\{ H_2 , H \}_{\mathfrak{so}(3)}=0$ for all Hamiltonian $H$;
hence $H_2$ is the \emph{Casimir} of the Poisson bracket (\ref{so(3)_Lie-Poisson})\,\cite{Casimir}.
% \footnote{The \emph{Casimir} $C$ is a (nontrivial) element such that $\{C, H \}=0$ for all $H$,
% i.e., it is a member of the \emph{center} of the Poisson algebra.
% }

\item[(ii)]
The generalized Boltzmann distribution (\ref{generalized_Boltzmann}) may be viewed as a standard Boltzmann distribution on a grand canonical ensemble with \emph{chemical potentials} $\beta_2/\beta_1, \cdots, \beta_{m}/\beta_1$ and \emph{charges} $H_2,\cdots,H_{m}$
($\beta_1$ is the inverse temperature and $H_1$ is the energy).
Or, one may remember the Boltzmann distribution of magnetized particles, where one may interpret  $\beta_2$ as the magnetic moment, and $H_2$ as the particle number.

\item[(iii)]
Comparing (i) and (ii), we may interpret the Casimir (one of the multiple Hamiltonians) is an \emph{adiabatic invariant} or an \emph{action variable}
(see \cite{YoshidaMahajan2014,Yoshida2016,YoshidaMorrison2022} for the argument of interpreting a Casimir as an adiabatic invariant).

\end{enumerate}

In these arguments, the additional Hamiltonians are assigned as specific quantities.  
A more ambitious interpretation of Nambu's proposal suggests arbitrariness to the choice of multiple Hamiltonians.
Then, one should impose on the bracket the so-called \emph{fundamental identity}\,\cite{Takhtajan1994}
that reads, for $m=2$,
\begin{eqnarray}
\{\{A, B, C\} ,H_1, H_2 \} &=& \{ \{A,  H_1, H_2 \}, B, C \} 
\nonumber \\
& & + \{ A, \{B,  H_1, H_2 \},  C \}+ \{ A, B, \{C,  H_1, H_2 \} \} .
\label{fundamental}
\end{eqnarray}
This is the requirement for the Nambu bracket to work as the generator of dynamics satisfying the derivation property (Leibniz law);  
if $\frac{\rmd}{\rmd t} G = \{ G, H_1, H_2 \}$,
 \begin{equation}
\frac{\rmd}{\rmd t} \{A, B, C\}  =  \{ \frac{\rmd}{\rmd t} A, B, C \} + \{  A, \frac{\rmd}{\rmd t}B, C \} +  \{  A, B, \frac{\rmd}{\rmd t}C \}
\label{fundamental-2}
\end{equation}
means (\ref{fundamental}).
This relation reduces into the Jacobi identity, 
if we specify the second Hamiltonian $S$ and put $\{ G , H, S\} =\{G, H \}_{S}$.
In fact, inserting $C=H_2=S$, (\ref{fundamental}) reads
\begin{equation}
\{\{A, B\}_S ,H_1 \}_S = \{ \{A, H_1 \}_S, B \}_S + \{ A,  \{B,  H_1 \}_S \}_S ,
\label{fundamental-reduced}
\end{equation}
which is nothing but the Jacobi identity.
The Poisson bracket $\{G, H \}_{S}$ has a Casimir $S$, because $\{G, S \}_{S}=\{ G, S, S \}=0$ ($\forall G$);
cf. the example (\ref{so(3)_Lie-Poisson-2}). 
Therefore, if one demands the Nambu bracket to satisfy the fundamental identity, 
it must reduce to a noncanonical Poisson bracket for every fixed $S$.
This is indeed a tall order (see Sec.\,\ref{sec:Lie-Poisson}).

Here, we do not explore the fundamental identity.
Instead, we consider the Nambu bracket for a special set of Hamiltonians, and study it as a noncanonical Poisson bracket with Casimirs. 
We discuss the relation among Lie algebras, Poisson brackets, and Casimirs.
Most of the contents may be known to specialists, but it may worth describing the relation of these notions in the perspective of Nambu dynamics.
One outcome is a natural correspondence principle for quantizing Nambu dynamics.

%%%%%%%%%%%%%%%%%%%%%%%%%%%%%%%%%%%%%%%%%%%%%%%%%%%%%%%%%%%%%%%%%%%%%%%%%%%%%%%%%%%%%%%%%%%
\section{Lie-Poisson bracket}
\label{sec:Lie-Poisson}

The example (\ref{Nambu_bracket}) of the Nambu bracket draws heavily on the specialty of the underlying $\mathfrak{so}(3)$ Lie algebra.
To see how we can generalize it, we elucidate the mathematical structure of this example in the perspective of Lie-Poisson algebra\,\cite{Morrison1998,MarsdenWeinstein1983}.

\subsection{Basic definitions}
\label{subsec:basic_definitions}

We prepare basic notation.
Let us denote by $[~,~]$ the bracket of a Lie algebra $\mathfrak{g}$.  
The natural coupling of the vector space $\mathfrak{g}$ and its dual $\mathfrak{g}^*$ is written as
$\langle \bi{x}, \bxi \rangle$ ($\bi{x}\in \mathfrak{g}, ~ \bxi\in\mathfrak{g}^* $).
We call $\mathfrak{g}$ the \emph{state space} (a member $\bi{x}$ of $\mathfrak{g}$ is a \emph{state vector}),
and $\mathfrak{g}^*$ the \emph{phase space}  (a member $\bxi$ of $\mathfrak{g}^*$ is an \emph{observable})\,\cite{covariant}.
% \footnote{
% In Sec.\,\ref{sec:introduction}, we did not distinguish the contravariant and covariant variables.
% In the following discussions, the duality of both vectors will be important;
% $\mathfrak{g}$ is the space of contravariant vectors (operators), while $\mathfrak{g}^*$ is the space of covariant vectors (observables).
% }
A general \emph{physical quantity} is a smooth function $ G(\bxi) \in C^\infty(\mathfrak{g}^*)$. 

By $\ad_{\bi{v}} \circ = [ \circ, \bi{v}] : \, \mathfrak{g} \rightarrow \mathfrak{g}$, we denote the adjoint representation of $\bi{v}\in \mathfrak{g}$.
Physically, the action of $\ad_{\bi{v}}$ on a state vector $\bi{x}\in \mathfrak{g}$ 
represents the infinitesimal dynamics:
\[
\dot{\bi{x}}= \ad_{\bi{v}} \bi{x} = [\bi{x}, \bi{v}] .
\]
Dual to  $\ad_{\bi{v}}$,
we define the coadjoint action $\ad^*_{\bi{v}} \circ = [\bi{v},\circ]^*:\, \mathfrak{g}^* \rightarrow \mathfrak{g}^*$,
where $[~,~]^*: \mathfrak{g}\times \mathfrak{g}^* \rightarrow \mathfrak{g}^*$ is defined by\,\cite{semi-simple}
% \footnote{
% For a semi-simple Lie algebra (like $\mathfrak{so}(3)$), we can formally evaluate $[\bi{x},\bi{y}]^* = [\bi{x},\bi{y}]$.
% This means that we may identify $\mathfrak{g}=\mathfrak{g}^*$ with an appropriate basis of $\mathfrak{g}^*$,
% and that the structure constants are \emph{fully antisymmetric}; for example, see \cite{YM2020}.
% We note that the general dual bracket $[~,~]^*$ is not necessarily a Lie bracket 
% (even $[\bxi,\bxi]^* =0$ may not hold).
% }
\begin{equation}
\langle [\bi{x},\bi{v}], \bxi \rangle = : \langle \bi{x}, [\bi{v}, \bxi]^* \rangle  .
\label{coadjoint}
\end{equation}
The left-hand side means that we observe the dynamics of a state vector $\bi{x}$
by measuring an observable $\bxi$.
The right-hand side is its translation into the change in the observable $\bxi$ (evaluated for a fixed state vector $\bi{x}$):
\[
\dot{\bxi}=\ad^*_{\bi{v}} \bxi = [\bi{v},\bxi]^*.
\]

By introducing a basis for each space,
let us examine the mutual relationship between $\mathfrak{g}$ and $\mathfrak{g}^*$ more explicitly.
When $\mathfrak{g}$ has a finite dimension $n$, we can define a basis $\{\bi{e}_1,\cdots,\bi{e}_n \}$
to represent $\bi{x} =  x^k \bi{e}_k$
(we invoke Einstein's summation rule of contraction).
On the other hand,  we provide $\mathfrak{g}^*$ with the dual basis $\{\bi{e}^1,\cdots,\bi{e}^n \}$ such that
$\langle \bi{e}_j, \bi{e}^k \rangle = \delta_{jk}$.
A complete system of measurements is given by $\Xi^k (\bi{x})= \langle \bi{x} , \bi{e}^k \rangle $
($k=1,\cdots,n$);
measuring every $x^k = \langle \bi{x} , \bi{e}^k \rangle$ for a state vector $\bi{x}$, 
we can identify it as $\bi{x}= x^k \bi{e}_k$.
Therefore, we may say that $\mathfrak{g}^*$ defines $\mathfrak{g}$ as $(\mathfrak{g}^*)^*$\,\cite{reflexive,topology_fs}.
% \footnote{
% This \emph{reflexive relation} is not trivial in infinite dimensions.}
% \footnote{
% In the case of infinite dimension vector spaces (function spaces), the \emph{topology} of $\mathfrak{g}^*$ determines the ``accuracy'' of measurement.
% For example, if we chose $\mathfrak{g}^*$ to be $C^\infty$-class, we obtain rather rough picture of phenomena in $\mathfrak{g}$ represented by generalized functions (distributions).
% If we change $\mathfrak{g}^*$ to $L^2$-class, the corresponding picture of $\mathfrak{g}$ becomes $L^2$-class functions.
% }
In fact, it is more legitimate to construct a theory by first defining $\mathfrak{g}^*$, since the description of a system depends on what we can measure.
For example, if we remove $\bi{e}^n$ from $\mathfrak{g}^*$, the component $x^n$ becomes invisible,
resulting in a reduced identification of $\bi{x}$ by only $x^j=\langle \bi{x}, \bi{e}^j\rangle$ ($j=1,\cdots, n-1$).

Changing the phase space $\mathfrak{g}^*$ will be the central issue of the following discussions.
Here we note that dropping $\bi{e}^n$ from $\mathfrak{g}^*$ may affect seriously in the representation of the dynamics in $\mathfrak{g}$;
for example, the modified dynamics may no longer be Hamiltonian, i.e., the corresponding bracket may fail to satisfy the Jacobi identity.

\subsection{General construction of Lie-Poisson brackets}
\label{subsec:general_Lie-Poisson}

The Poisson bracket is a Lie bracket defined on a Banach ring, which satisfies the Leibniz law such as $\{ FG, H \}=F \{G,H \}+ G \{F,H \}$ in addition to the usual axioms of Lie brackets.
There is a systematic way of formulating a Poisson bracket from a Lie bracket; we use the relation (\ref{coadjoint}).
Such a bracket is called a \emph{Lie-Poisson bracket}.

For $G(\bxi)\in C^\infty(\mathfrak{g}^*)$, we define its \emph{gradient} $\partial_{\bxi} G \in \mathfrak{g}$ by
(a perturbation in $u$ is denoted by $\tilde{u}$)
\begin{equation}
\tilde{G} = G(\bxi+\epsilon\tilde{\bxi}) - G(\bxi) 
= \epsilon \langle \partial_{\bxi} G, \tilde{\bxi} \rangle + O(\epsilon^2)
\quad (\forall \tilde{\bxi}\in \mathfrak{g}^*).
\label{gradient}
\end{equation}
The phase space $\mathfrak{g}^*$ is made a Poisson manifold by endowing it with
\begin{equation}
\{ G, H \} = \langle  [\partial_{\bxi} G ,\partial_{\bxi} H], \bxi \rangle
= \langle  \partial_{\bxi} G , [\partial_{\bxi} H, \bxi]^* \rangle .
\label{Lie-Poisson_definition}
\end{equation}
Because of this construction, $\{ ~,~\}$ inherits 
bilinearity, anti-symmetry, and the Jacobi identity from that of $[~,~ ]$. 
The Leibniz property is implemented by the derivation $\partial_{\bxi}$, so $\{~.~\}$ is a Poisson bracket.

Denoting
\begin{equation}
J(\bxi) \circ  = [\, \circ \, , \bxi]^* \quad : \, \mathfrak{g} \rightarrow \mathfrak{g}^*,
\label{Poisson_operator}
\end{equation}
which we call a \emph{Poisson matrix} (or \emph{Poisson operator}, particularly if $\mathfrak{g}$ is an infinite-dimensional space).
Notice that $\bi{x}\in\mathfrak{g}$ is a contravariant vector, while $\bxi\in\mathfrak{g}^*$ is a covariant vector.
Hence, $J$ is a 2-form.

we may write (\ref{Lie-Poisson_definition}) as
\begin{equation}
\{ G,H \} = \langle \partial_{\bxi} G , J(\bxi) \partial_{\bxi} H \rangle .
\label{Lie-Poisson-3}
\end{equation}

The evolution of a physical quantity $G \in C^\infty(\mathfrak{g}^*)$, generated by a \emph{Hamiltonian} $H \in C^\infty(\mathfrak{g}^*)$ is dictated by
\begin{equation}
\dot{G} = \{ G, H \} = \langle  [\partial_{\bxi} G ,\partial_{\bxi} H], \bxi \rangle ,
\label{Hamilton's_equation_general}
\end{equation}
which reads Hamilton's equation for $G(\bxi) = \xi_j = \langle \bi{e}_j , \bxi \rangle$:
\begin{equation}
\dot{\xi_j} = \{ \xi_j , H \} = (J(\bxi) \partial_{\bxi} H )_j
\quad (j=1,\cdots,n).
\label{Hamilton's_equation_general-component}
\end{equation}

% \bigskip
\begin{example}[Heisenberg algebra and canonical Poisson bracket]
\label{ex:Heisenberg}
\normalfont
As the primary example, let us consider the Heisenberg algebra characterized by 
\begin{equation}
[\bi{e}_1,\bi{e}_2]=0,~ [\bi{e}_1,\bi{e}_3]=0,~[\bi{e}_2,\bi{e}_3]=\bi{e}_1.
\label{Heisenberg}
\end{equation}
Using this Lie bracket, we obtain
(denoting $\bxi=(r, q, p )$)
\begin{equation}
\{ G, H \} = \langle [\partial_{\bxi} G, \partial_{\bxi} H ], \bxi \rangle = \langle \partial_{\bxi} G, J \partial_{\bxi} H \rangle,
\quad 
J = \left( \begin{array}{ccc}
0 & 0 & 0 \\
0 & 0& r \\
0 & -r& 0 
\end{array} \right). 
\label{Heisenberg-Poisson}
\end{equation}
Evidently, $\{ r , H \}=0$ ($\forall H$), i.e., $r$ is the Casimir.
The $r=$ constant surface is the symplectic manifold, on which $(q,~p)$ is the canonical coordinate. 
Putting $r=1$, we obtain the canonical Poisson bracket
\begin{equation}
\{ G, H \} =  \partial_q G \partial_p H - \partial_p G \partial_q H  .
\label{Heisenberg-Poisson-1}
\end{equation}
\end{example}

\subsection{$\mathfrak{so}(3)$ Lie-Poisson bracket}
\label{subsec:so(3)_Lie-Poisson}

As illuminated by (\ref{Nambu_bracket}), the prototype Nambu bracket is related to the $\mathfrak{so}(3)$ algebra,
which is defined by
\begin{equation}
[\bi{e}_1,\bi{e}_2]_{\mathfrak{so}(3)} =\bi{e}_3,~ 
[\bi{e}_1,\bi{e}_3]_{\mathfrak{so}(3)} =-\bi{e}_2,~
[\bi{e}_2,\bi{e}_3]_{\mathfrak{so}(3)} =\bi{e}_1 .
\label{so(3)}
\end{equation}
In vector analysis notation, we may write, for $\bi{x}=x^j\bi{e}_j$ and $\bi{y}=y^j\bi{e}_j$, $[\bi{x}, \bi{y}]_{\mathfrak{so}(3)} =\bi{x}\times\bi{y}$.
The corresponding Lie-Poisson bracket is (denoting $\bxi= (\xi_1, \xi_2, \xi_3 )$)
\begin{eqnarray}
\{ G , H \}_{\mathfrak{so}(3)} &=& \langle [\partial_{\bxi}G, \partial_{\bxi}H ]_{\mathfrak{so}(3)}, \bxi \rangle 
= \langle \partial_{\bxi} G, J_{\mathfrak{so}(3)} \partial_{\bxi} H \rangle,
\nonumber \\
J_{\mathfrak{so}(3)} &=& \bxi\times = \left( \begin{array}{ccc}
0 & \xi_3 & -\xi_2 \\
-\xi_3 & 0 & \xi_1 \\
\xi_2 & \xi_1& 0 
\end{array} \right). 
% J_{\mathfrak{so}(3)} = \left( \begin{array}{ccc}
% 0 & p & -q \\
% -p & 0 & r \\
% q & r& 0 
% \end{array} \right). 
\label{so(3)-Poisson}
\end{eqnarray}
Evidently, $C = \frac{1}{2} |\bxi|^2$ is the Casimir.

\subsection{Variety of the 3-dimensional Nambu brackets}
\label{subsec:varity_3D}

The 3-dimensional Lie algebras are classified into nine Bianchi types (for example, see \cite{Ellis,Ryan});
the Heisenberg algebra is type-II (Example\,\ref{ex:Heisenberg}), and $\mathfrak{so}(3)$ is type-IX.
See \cite{YMT2017} for the total list of the 3-dimensional Lie-Poisson brackets and the corresponding Casimirs.
Therefore, we have nine types of 3-dimensional Nambu algebras.

Notice that the ``second Hamiltonian'' is restricted to the Casimirs.
Let us see what happens when we chose a different function $H_2 \in C^\infty(\mathfrak{g}^*)$ for the second Hamiltonian in the $\mathfrak{so}(3)$ 
Nambu bracket (\ref{Nambu_bracket}).
Suppose that $H_2$ is a quadratic function, and write 
\[
\partial_{\bxi} H_2 = M \bxi
\quad (M \in \mathrm{Hom}(\mathfrak{g}^*, \mathfrak{g}) ).
\]
For $\mathfrak{so}(3)$, we may identify $\mathfrak{g}^* = \mathfrak{g}$.  
Hence, (\ref{Nambu_bracket}) reads
\begin{equation}
\{G, H, H_2\} =  \nabla G \cdot (\nabla H)\times(M \bxi) 
=
\langle M^{\mathrm{T}}[\partial_{\bxi}G, \partial_{\bxi}H ]_{\mathfrak{so}(3)}, \bxi \rangle.
\label{Nambu_bracket_deformed}
\end{equation}
For this to satisfy the Jacobi identity, 
\begin{equation}
[~,~]_M := M^{\mathrm{T}} [~, ~]_{\mathfrak{so}(3)}
\label{deformed_bracket}
\end{equation}
must be a Lie bracket.
Therefore, changing the second Hamiltonian $H_2$ in the $\mathfrak{so}(3)$ Nambu bracket is equivalent to changing the underlying Lie algebra.
By Theorem 1 of \cite{YM2020}, 
every possible 3-dimensional Lie bracket is derived from the \emph{mother bracket}
$[~,~]_{\mathfrak{so}(3)}$ by the \emph{deformation} $[~,~]_M=M^{\mathrm{T}} [~, ~]_{\mathfrak{so}(3)}$ with some $M$;
hence, we only have nine possibilities
% \footnote{
% As far as $M$ is symmetric, it can be arbitrary.  Bianchi classifies them into six different types by geometrical reason; they are called class-A.
% Nonsymmetric $M$ is rather limited for $[~,~]_M$ to satisfy the Jacobi identity, and only six types are allowed; they are called class-B
% (among them, three types include parameters by which they degenerate into the corresponding class-A algebras, and hence, Bianchi types are classified into nine types): see Table VI of \cite{YM2020}.}\footnote{
% Only the $\mathfrak{so}(3)$ algebra is the \emph{mother} of all other 3-dimensional Lie algebras.
% If we chose some class-B algebra, and apply $M$ of rank $<3$ (i.e., drop some $\bi{e}^j$ from the observables), the resultant $[~,~]_M$ is no longer a Lie bracket. 
% }
in selecting the second Hamiltonian for the 3-dimensional Nambu bracket (if we permit that the first Hamiltonian is free)\,\cite{symmetric,mother}.

%%%%%%%%%%%%%%%%%%%%%%%%%%%%%%%%%%%%%%%%%%%%%%%%%%%%%%%%%%%
\section{Reduction, Clebash paramterization and gauge group}
\label{sec:reduction}

In the forgoing section, we examined possible varieties of Nambu brackets in the framework of Lie-Poisson algebras.
The key was to deform the observables ($\in \mathfrak{g}^*$) by a homomorphism (linear map $M$).
Here we consider a different type of deformation (in fact, a reduction) which is a bilinear map on $\mathfrak{g}^*$.

%%%%%%%%%%%

The fundamental identity (\ref{fundamental}) implies that a Nambu bracket with multiple Hamiltonians ($H_1,\cdots,H_m$) reduces into a Poisson bracket, if we chose one (say $H_1$) for the usual Hamiltonian and fix all others ($H_2,\cdots,H_m$) to some given functions; see (\ref{fundamental-reduced}).
Then, the fixed ones can be interpreted as Casimirs of the Poisson bracket $\{~,~\}$, i.e.,
$\{ G, H_j\} =0$ ($\forall G$; $j=2,\cdots,n$).
Writing 
\[
\{ G, H \} = \langle \partial_{\bxi} G, J  \partial_{\bxi} H \rangle ,
\]
a Casimir is the integral of the kernel of the Poisson operator $J$,
i.e., a Casimir $C$ is a function such that $\partial_{\bxi} C \in \mathrm{Ker} J $.
A Poisson bracket that has non-zero nullity is said \emph{noncanonical}\,\cite{Morrison1998}.
We note that the kernel of $J$ is not necessarily integrable (i.e., the symplectic foliation is, in general, an immersion, not embedding).
However, if a Casimir exists, its level-sets include the symplectic leaves.

The aim of this section is to characterize a noncanonical Poisson algebra as a reduction (subalgebra) of some canonical Poisson algebra.
To put it another way, we can extend the phase space to subsume the noncanonical Poisson system as a submanifold of the extended canonical system. 
The Casimirs help to find canonical variables to inflate the phase space.

\subsection{Angular momentum reduction to the $\mathfrak{so}(3)$ Lie-Poisson algebra}
\label{subsec:so(3)}
We start again with the noncanonical $\mathfrak{so}(3)$ Lie-Poisson algebra, the principal example of the Nambu bracket.
Changing the perspective, however, we derive it from a higher-dimensional canonical Poisson algebra.

As well-known\,\cite{Marsden}, $\{ G, H \}_{\mathfrak{so}(3)}$, given by (\ref{so(3)-Poisson}),
can be viewed as the ``angular momentum reduction'' of the canonical 6-dimensional symplectic algebra 
% $M_6=\mathbb{R}^6$\,\cite{Marsden}.
$\mathfrak{sp}(6,\mathbb{R})$.
The canonical Poisson bracket $\{G,H \}_{\mathfrak{sp}(6)}$ is made consistent to our notation by remembering Example\,\ref{ex:Heisenberg}.
The dual space $\mathfrak{sp}(6,\mathbb{R})^* \cong \mathbb{R}^6 $ of $\mathfrak{sp}(6,\mathbb{R})$ is the phase space.
Let $\bi{z}=(\bi{q},\bi{p})^{\mathrm{T}}$, consisting of the position vector $\bi{q}\in \mathbb{R}^3$ and the momentum vector $\bi{p}\in  \mathbb{R}^3$, be the canonical coordinates of the phase space.
Then, the canonical bracket is written as
\begin{equation}
\{ G, H \}_{\mathfrak{sp}(6)} = \langle \partial_{\bi{z}} G, J_c \partial_{\bi{z}} H \rangle,
\quad 
J_c = \left( \begin{array}{cc}
 0 & I  \\
-I & 0 \\
\end{array} \right),
\label{canonical_bracket}
\end{equation}
which is the direct product of the Heisenberg Lie-Poisson bracket\,cite{Heisenberg}; see (\ref{Heisenberg-Poisson}).
% \footnote{
% As seen in the general construction of the Lie-Poisson bracket (Sec.\,\ref{subsec:general_Lie-Poisson}),
% the Poisson operator $J$ is a linear function of the phase space coordinates.
% In (\ref{Heisenberg-Poisson}),  $J$ depends on $r$, but $r$ is a Casimir, so it is constant on each symplectic leaf.
% }

We consider a subsystem in which all physical quantities only depend on the angular momentum 
\begin{equation}
\bxi=\bi{q}\times\bi{p}.
\label{angular_momentum}
\end{equation}
Euler top is such an example.
Then, the canonical bracket 
$\{ G, H \}_{\mathfrak{sp}(6)}$,
when evaluated for $G(\bxi)$ and $H(\bxi)$, reduces to $\{ G, H \}_{\mathfrak{so}(3)}$.
In fact, for every smooth function $F(\bxi)$, 
we can evaluate the gradient by
\[
\tilde{F} = \langle \partial_{\bi{q}} F, \tilde{{\bi{q}}} \rangle +
 \langle \partial_{\bi{p}} F, \tilde{{\bi{p}}}  \rangle
=  \langle \partial_{\bxi} F, \tilde{{\bxi}} \rangle\, .
\]
Inserting $ \tilde{{\bxi}} = \tilde{\bi{q}}\times\bi{p} + \bi{q}\times\tilde{\bi{p}}$, we obtain
\[
\partial_{\bi{q}} F = \bi{p}\times\partial_{\bxi} F, \quad 
\partial_{\bi{p}} F = -\bi{q}\times\partial_{\bxi} F .
\]
Using these relations, the canonical bracket $\{G, H\}_{\mathfrak{sp}(6)}$ evaluates, for every $G(\bxi)$ and $H(\bxi)$, as
\begin{equation}
\{G(\bxi), H(\bxi) \}_{\mathfrak{sp}(6)} = 
 \langle \partial_{\bxi} G, \partial_{\bxi} H \times \bxi  \rangle
=  \{G(\bxi), H(\bxi) \}_{\mathfrak{so}(3)} .
\label{so(3)_canonicalization}
\end{equation}

\subsection{Canonicalization and gauge symmetry}
\label{subsec:gauge}

When we interpret (\ref{so(3)_canonicalization}) from the opposite view point,
the noncanonical Poisson brackt $\{G, H \}_{\mathfrak{so}(3)}$ can be \emph{canonicalized} as $\{G, H\}_{\mathfrak{sp}(6)}$ in the larger phase space $\mathfrak{sp}(6,\mathbb{R})^*$.
For a given $\bxi$, representing it as $\bxi = \bi{p}\times\bi{q}$ by two ``potentials'' $\bi{p}$ and $\bi{q}$ is said \emph{Clebsch parameterization}.
% \footnote{The original idea of Clebsch parameterization is described in Sec.\, \ref{subsec:compressible_fluid_Clebsch}.}

The Casimir is a characteristic of noncanonical Poisson algebras.
When canonicalized, the Poisson operator is regular, so there is no physical quantity that is invariant against arbitrary Hamiltonian flows.
In the canonicalized larger Poisson algebra (Hamiltonian system), the invariance of the original Casimir $C$ is due to the specific \emph{symmetry} of the Hamiltonians allowed in the original noncanonical system,
which indeed means the \emph{gauge symmetry} of the Clebsch parameterization.

Before examining the geometrical meaning of the gauge symmetry of the foregoing $\mathfrak{so}(3)$ example,
we formulate the basic relations.
Let $\bxi\in \mathfrak{g}^*_{nc}$ be the original observable of the noncanonical Poisson manifold endowed with the bracket $\{~,~\}_{nc}$,
and $\bxi= \bxi (\bi{z})$ be the Clebsch parameterization by the potential variables $\bi{z} \in  \mathfrak{g}^*_{c}$;
on the canonicalized phase space $\mathfrak{g}^*_{c}$, we have a canonical Poisson bracket $\{~,~\}_c$.
The \emph{reduction} means 
\[
\{ G(\bxi(\bi{z})), H(\bxi(\bi{z})) \}_c = \{ G(\bxi), H(\bxi) \}_{nc} .
\]
Remember (\ref{so(3)_canonicalization}) for an example.
Suppose that $C(\bxi)$ is a Casimir of $\{~,~\}_{nc}$.
Then, $\{ F(\bxi(\bi{z})), C \}_{c}=0$ ($\forall F(\bxi(\bi{z}))$), which means
$\{ \xi_j , C \}_c=0$ ($\forall j$).

There is a function $\theta(\bi{z})$ such that $\{ \theta(\bi{z}),  C(\bxi(\bi{z})) \}_{c} =1$ at least locally in $\mathfrak{g}^*_{c}$.
Then, $\{ \circ, C \}_{c} = \partial_\theta$,
i.e., the gauge symmetry means the independence of $\bxi(\bi{z})$ on $\theta$.

Now we do practice with the $\mathfrak{so}(3)$ example and unearth the gauge parameter $\theta$.
The level-set of the Casimir $\frac{1}{2}|\bxi|^2$ is the symplectic leaf,
which is a sphere in the 3-dimensional space of the angular momentum $\bxi$.
Transforming $\bxi$ to the local coordinates $\bxi'=(C,  \phi,\xi_1)$, where
$\phi = \tan^{-1} \xi_2/\xi_3$,
$J_{\mathfrak{so}(3)} $ is rewritten as
\begin{equation}
J_{\mathfrak{so}(3)}' = 
\left( \begin{array}{ccc}
0 & 0 & 0 \\
0 & 0 & 1 \\
0 & -1 &  0
\end{array} \right) ,
\label{3D-J'}
\end{equation}
in which the kernel is apparent.
Separating the kernel $\xi'_1 = C$ from the 3-dimensional phase space of $\bxi$, we may canonicalize the $\mathfrak{so}(3)$ Poisson operator as
\begin{equation}
J_{\mathfrak{so}(3)}' = 
\left( \begin{array}{c:cc}
0 & 0 & 0 \\
\hdashline
0 & 0 & 1 \\
0 & -1 &  0
\end{array} \right) 
~\mapsto ~
J'_2 = 
\left( \begin{array}{cc}
 0 & 1 \\
-1 &  0
\end{array} \right) ,
\label{3D-J'-canonize}
\end{equation}
which dictates the symplectic structure of the leaf $C=$ constant.

There is the other way of canonicalization; we inflate the phase space to 4-dimension by adding a conjugate variable $\theta$ such that $\{\theta, C\}_c = 1$.
Let 
\[
\theta = \frac{1}{2|\bxi|} \tan^{-1} \left( \frac{ ( \bxi\times\bi{q})_j}{ q_j |\bxi| } \right) 
% \label{conjugate-theta}
\]
(we choose the coordinate $q_j\neq0$), and inflate $\bxi' =(C,  \phi, \xi_1) \mapsto  \widetilde{\bxi'} =(\theta,C, \phi, \xi_1) $ to canonicalize
\begin{equation}
J_{\mathfrak{so}(3)}' = 
\left( \begin{array}{ccc}
0 & 0 & 0 \\
0 & 0 & 1 \\
0 & -1 &  0
\end{array} \right) 
~\mapsto ~
J'_4 = 
\left( \begin{array}{c:ccc}
0 & 1 & 0 & 0 \\
\hdashline
-1& 0 & 0 & 0 \\
0 & 0 & 0 & 1 \\
0 & 0 & -1 &  0
\end{array} \right)  .
\label{3D-J'-canonize-2}
\end{equation}
The  action $\{ \circ, C \}_4 = \partial_\theta$
yields the co-rotation of $\bi{q}$ and $\bi{p}$ around the axis $\bxi$,
which evidently conserves $\bxi=\bi{p}\times\bi{q}$.
% Hence, the adjoint action $\{ \circ, C \}_4 $ generates the \emph{gauge group} of $\bxi=\bi{q}\times\bi{p}$.

% This example suggests the general method for removing nullity pertinent to Casimirs;
% we add a conjugate variable to change a Casimir (invariant due to the degeneracy of the Poisson operator $J$) to be variable. 

\subsection{Spin reduction to the $\mathfrak{su}(2)$ Lie-Poisson algebra}
\label{subsec:su(2)}
By $\mathfrak{so}(3) \cong \mathfrak{su}(2)$, we may formulate an alternative reduction to an equivalent noncanonical Poisson algebra,
or an alternative Clebsch parameterization;
instead of the angular momentum in $\mathbb{R}^6$, we consider the spin in $\mathbb{C}^2$.

We consider the 2 dimensional complex phase space, and write
\[
\bi{z}= \left( \begin{array}{c} {z}_1 \\ {z}_2 \end{array} \right)  \in \mathbb{C}^2 ,
\quad 
\left\{ \begin{array}{l}
z_1 = q_1 + \rmi p_1, \\
z_2 = q_2 + \rmi p_2.
\end{array} \right.
\]
By the correspondence $\bi{z} \Leftrightarrow \bi{Z} = (\Re \bi{z}, \Im \bi{z} )^{\mathrm{T}} =  (q_1,q_2,p_1,p_2)^{\mathrm{T}} $,
we define the canonical Poisson bracket 
\[
\{ G(\bi{Z}), H(\bi{Z}) \}_4 = \langle \partial_{\bi{Z}} G(\bi{Z}), J_c \partial_{\bi{Z}} H(\bi{Z}) \rangle .
\]

We consider the reduction of $\bi{Z}$ (or $\bi{z}$) to the 3-dimensional spin parameters.
Using the Pauli matrices 
\[
\sigma_1 = \left( \begin{array}{cc}
0 & 1 \\
1 & 0
\end{array} \right),
\quad 
\sigma_2 = \left( \begin{array}{cc}
0 & -i \\
i  & 0
\end{array} \right),
\quad 
\sigma_3 = \left( \begin{array}{cc}
1 & 0 \\
0 & -1
\end{array} \right)
\]
we define
\begin{equation}
\xi_j = \frac{1}{4} \bi{z}^* \sigma_j \bi{z}
\quad (j=1,2,3).
\label{Cayley-Klein-2}
\end{equation}
In terms of $\bi{Z}$, we may write
\begin{equation}
\bxi = \left( \begin{array}{c}
\xi_1 \\ \xi_2 \\ \xi_3
\end{array} \right)
=\frac{1}{2} \left( \begin{array}{c}
q_1 q_2+p_1 p_2 \\
q_1 p_2 - q_2 p_1 \\
(q_1^2 + p_1^2 - q_2^2 - p_2^2)/2
\end{array} \right) ,
\label{Cayley-Klein}
\end{equation}
which we call the spin parameterization (also known as \emph{Cayley-Klein parameterization}).
Inserting these parameters, we find
\[
\{ \xi_j, \xi_k \}_4 = \epsilon_{jk\ell} \, \xi_\ell .
\]
Therefore, we obtain the reduction
\[
\{G(\bxi(\bi{Z})), H(\bxi(\bi{Z})) \}_4
=  \{G(\bxi), H(\bxi) \}_{\mathfrak{so}(3)} .
\]
Conversely, the spin representation (\ref{Cayley-Klein}) is an alternative Clebsch parameterization of the $\mathfrak{so}(3)$ Lie-Poisson algebra, which is a more economical canonization by the 4-dimension parameters, in comparison to the previous angular momentum representation (\ref{angular_momentum}) by the 6-dimension parameters.

We easily find 
\begin{equation}
C = 2 |\bxi| =\frac{1}{2} (|z_1|^2 + |z_2|^2) 
% = \frac{1}{2} (q_1^2 + q_2^2 + p_1^2 + p_2^2)
\label{su(2)-Casimir}
\end{equation}
is the Casimir, and its conjugate variable $\theta$ (such that $\{\theta, C \}_4 =1$) is
\[
\theta = \frac{1}{2} (\mathrm{Arg}\, z_1 + \mathrm{Arg}\, z_2)
= \frac{1}{2} \left[ \arctan \left( \frac{q_1}{p_1}\right) +  \arctan \left( \frac{q_2}{p_2}\right) \right].
\]
The gauge transformation reads
\[
\{ C, \bi{Z} \}_4 \Leftrightarrow  i \left( \begin{array}{cc}
1 & 0 \\
0 & 1 
\end{array} \right) \bi{z} ,
\]
which means the $U(1)$ gauge transformation.
Therefore, the spin parameters, together with the $U(1)$ gauge, span the 4 dimensional canonical Poisson manifold.

%%%%%%%%%%%%%%%%%%%%%%%%%%%%%%%%%%%%%%%%%%%%%%%%%%%%%%%%%%%%%%%%%%%%%%%%%%%%%%%%%%%%%%%%%%%%%%%%%%%%%%%%%%%%%%
%%%%%%%%%%%%%%%%%%%%%%%%%%%%%%%%%%%%%%%%%%%%%%%%%%%%%%%%%%%%%%%%%%%%%%%%%%%%%%%%%%%%%%%%%%%%%%%%%%%%%%%%%%%%%%
%%%%%%%%%%%%%%%%%%%%%%%%%%%%%%%%%%%%%%%%%%%%%%%%%%%%%%%%%%%%%%%%%%%%%%%%%%%%%%%%%%%%%%%%%%%%%%%%%%%%%%%%%%%%%%
\section{Vortex dynamics as Nambu mechanics}
\label{sec:vortex_dynamics}

As an example of classical field theories, we invoke the vortex equation of ideal (inviscid and incompressible) fluid, and formulate it as a noncanonical Hamiltonian mechanics.
The \emph{helicity} plays the role of the second Hamiltonian, if we view the system as a Nambu dynamics\,\cite{Fukumoto2009}.
% \footnote{
% The analogy of the vortex dynamics and Nambu dynamics was pointed out by Fukumoto\,\cite{Fukumoto2009}.}

\subsection{Basic equation}
\label{subsec:vortex_equation}
We start by reviewing the basic equation of vortex dynamics.
Let  be the base space; 
To avoid compexity pertinent to boundary conditions, we consider 3-torus $\Omega=\mathbb{T}^3$ for the domain hosting inviscid and incompressible fluid; the coordinate will be denoted by $\bi{x}$.
The fluid velocity, denoted by $\bi{v}$, is a 3-vector field on $\Omega$. 
The vorticity is $\bomega=\nabla\times\bi{v}$.
The Euler equation of incompressible flow ($\nabla\cdot\bi{v}=0$) is,
putting the constant mass density $\rho=1$ and the pressure $P$,
\begin{equation}
\partial_t \bi{v} = -(\bi{v}\cdot\nabla) \bi{v} - \nabla P .
\label{Euler}
\end{equation}
Operating curl on the both sides of (\ref{Euler}) yields the vortex equation
\begin{equation}
\partial_t \bomega = \nabla\times (\bi{v}\times\bomega ) .
\label{vortex_eq}
\end{equation}

If we can evaluate $\bi{v} = \mathrm{curl}^{-1} \bomega$ with a unique inverse $\mathrm{curl}^{-1} $ of the curl,
(\ref{vortex_eq}) reads an autonomous equation governing $\bomega$:
\begin{equation}
\partial_t \bomega = \nabla\times ((\mathrm{curl}^{-1} \bomega)\times\bomega ) .
\label{vortex_eq-2}
\end{equation}
To define $\mathrm{curl}^{-1} $, we have to identify an appropriate function space of $\bi{v}$.
Let us decompose $L^2(\Omega)$ as
\[
L^2(\Omega) = L^2_\Sigma(\Omega) \oplus \mathrm{Ker (curl)}.
\]
Explicitly, $L^2_\Sigma(\Omega)$ is the totality of divergence-free and flux-free ($\int_\Sigma \bi{v}\cdot\rmd^2\bi{s} =0$ for every cross-section of $\Omega$) vector fields in $\Omega=\mathbb{T}^3$\,\cite{BC}.
% \footnote{In the case of a general 3-dimensional domain $\Omega$, $\bi{v}\in L^2_\Sigma(\Omega)$ also satisfies $\bi{n}\cdot\bi{v}=0$ on the boundary $\partial\Omega$.
% }
The reduced phase space is
\begin{equation}
V_{\bomega} = \{ \bi{v}\in L^2_\Sigma(\Omega) ;\, \nabla\times\bi{v} \in   L^2_\Sigma(\Omega) \} ,
\label{domain_curl}
\end{equation}
on which the $\mathrm{curl}$ is a self-adjoint operator with the compact inverse $\mathrm{curl}^{-1} $\,\cite{YG}.
We will denote by $\mathrm{curl}$ the self-adjoint curl operator $V_{\bomega} \rightarrow  L^2_\Sigma(\Omega) $.

% \footnote{
% We invoke the theory of self-adjoint curl operator given by \ref{YG}.
% The essential part of the theory is to separate the gradient fields, as well as harmonic fields, from the domain and range of the curl operator.
% For the incompressible flow $\bi{v}$ in $\Omega=\mathbb{T}^3$, we have to assume that $\bi{v}$ has no net flux though every cross-section of $\Omega$.
% Then, $\mathrm{curl}^{-1} $ is uniquly determined as a self-adjoint compact operator;
% see also \cite{YM2020} for the direct application to the present context.
% }

\subsection{Hamiltonian form of the vortex equation}
\label{subsec:Hamiltonian_vortex_equation}
To write the vortex equation in a Hamiltonian form, we consider a formal Poisson operator 
\begin{equation}
J_{\bomega} \, \circ = \mathrm{curl}\, [(\mathrm{curl}\, \circ) \times \bomega ],
\label{vortex_Poisson_naive}
\end{equation}
and define a bracket (for $G, H \in C^\infty(V_{\bomega} )$)
\begin{equation}
\dpoisson{G(\bomega)}{H(\bomega}_{\bomega} = \dangle{\partial_{\bomega}G}{J_{\bomega} \partial_{\bomega} H} ,
\label{vortex_Poisson_bracket_naive}
\end{equation}
where $\dangle{\bi{a}}{\bi{b}} = \int_\Omega \bi{a}\cdot\bi{b} \,\rmd^3x$
(we use double brackets for field variables);
see Remark\,\ref{remark:duality}.
By formal calculations, we can show that $\dpoisson{~}{~}_{\bomega}$ satisies the requirements of Poisson brackets.
The Hamiltonian of the fluid is
\begin{equation}
H(\bomega) = \frac{1}{2} \int_\Omega |\mathrm{curl}^{-1} \bomega|^2 \, \rmd^3 x .
\label{vortex_Hamiltonian}
\end{equation}
Using the self-adjointness of $\mathrm{curl}^{-1}$\,\cite{YG}, we obtain $\partial_{\bomega} H = \mathrm{curl}^{-2}\bomega$.
Hamilton's equation\,\cite{Hamilton_eq}
% \footnote{
% Comparing $\dot{G}=\dangle{\partial_{\bomega} G }{\dot{\bomega}}$ and
% $\dot{G}=\dpoisson{G}{H}_{\bomega} = \dangle{\partial_{\bomega} G}{J_{\bomega} \partial_{\bomega} H}$
% ($\forall G(\bomega)$), 
% we obtain (\ref{vortex_eq-Hamiltonian}) as an evolution equation in the topology of $L^2(\Omega)$.
% }
reproduces the vortex equation (\ref{vortex_eq-2}):
\begin{equation}
\dot{\bomega }=  \mathrm{curl}\, [(\mathrm{curl}^{-1}\bomega) \times \bomega ].
\label{vortex_eq-Hamiltonian}
\end{equation}
Evidently, 
\begin{equation}
C = \frac{1}{2} \dangle{\mathrm{curl}^{-1} \bomega}{ \bomega }
\label{helicity_of_vorticity}
\end{equation}
is the Casimir of the noncanonical bracket $\dpoisson{~}{~}_{\bomega}$,
which is known as the \emph{helicity}.

\begin{remark}[duality]
\label{remark:duality}
\normalfont
Here we evaluate $\dangle{~}{~}$ as the usual $L^2$ inner product.
In the later discussion (Sec.\,\ref{subsec:Clebsch_vorticity}), however, we invoke the differential geometrical definition of inner products.
We identify the vorticity $\bomega$ as a 2-form, and then, $\partial_{\bomega} F(\bomega)$ is a 2-form;
we evaluate
\[
\tilde{F}(\bomega) = \dangle{\partial_{\bomega} F(\bomega)}{\tilde{\bomega}}
= \int_\Omega \partial_{\bomega} F(\bomega) \wedge * \tilde{\bomega}
= \int_\Omega \langle (\partial_{\bomega} F(\bomega) )^\dagger , \tilde{\bomega} \rangle \rmd^3 x ,
\]
where $*$ is the Hodge star operator,  $\rmd^3 x$ is the volume 3-form, $(\partial_{\bomega} F(\bomega) )^\dagger = (* \partial_{\bomega} F(\bomega))/\rmd^3x$ is the 2-vector being the dual of the 2-form $\partial_{\bomega} F(\bomega)$, and
$\langle~,~\rangle$ is the pairing of the contravariant and covariant tensors.
In general, for a $p$-form $\bxi$ phase space element, $\partial_{\bxi} F(\bxi)$ is evaluated as a $p$-form,
and its dual  $(\partial_{\bxi} F(\bxi))^\dagger$ is a $p$-vector.
\end{remark}

\subsection{Vector bundle of $\mathfrak{so}(3)$ fibers}
\label{subsec:bundle}
Now we elucidate the relation between the $\mathfrak{so}(3)$ Lie-Poisson bracket (\ref{so(3)-Poisson})
and the vortex bracket (\ref{vortex_Poisson_bracket_naive}),
by which we may regard the vortex system as an example of Nambu dynamics.

On the base space $\Omega =\mathbb{T}^3$, we consider the vector bundle $E_{\Omega\times\mathfrak{so}(3)}$ that consists of fibers of the $\mathfrak{so}(3)$ algebra;
each fiber has the Lie bracket 
\[
[\bi{a}, \bi{b} ]_{\mathfrak{so}(3)} = \bi{a}\times \bi{b} 
\quad (\bi{a}, \bi{b} \in \mathbb{R}^3).
\]
A smooth cross section of $ E_{\Omega\times\mathfrak{so}(3)}$ is denoted by $\bi{v}(\bi{x})$ ($\bi{x}\in\Omega$).
The totality of smooth cross sections constitutes the function space $\mathfrak{V}$ of smooth 3-vector functions on $\bi{x}\in \Omega$,
which is endowed with a Lie bracket
\[
\dbracket{\bi{v}(\bi{x})}{\bi{w}(\bi{x})}_{\mathfrak{so}(3)} =   \bi{v}(\bi{x})\times \bi{w}(\bi{x}) ,
\quad (\bi{x}\in\Omega).
\]
The $L^2$-completion of $\mathfrak{V}$ is denoted by ${V}$.
Evidently, $\dbracket{~}{~}_{\mathfrak{so}(3)}^* =  \dbracket{~}{~}_{\mathfrak{so}(3)}$.

The field-theory version of the $\mathfrak{so}(3)$ Lie-Poisson bracket is
\begin{equation}
\dpoisson{G}{H}_{\mathfrak{so}(3)} := \dangle{ \dbracket{\partial_{\bi{v}} G}{\partial_{\bi{v}}H}_{\mathfrak{so}(3)} }{\bi{v} }
= \dangle{ \partial_{\bi{v}} G}{ \dbracket{\partial_{\bi{v}}H}{\bi{v}}_{\mathfrak{so}(3)}^* } ,
\label{SO(3)_of_fields}
\end{equation}
and the corresponding Poisson operator is
\begin{equation}
\mathcal{J}_{\mathfrak{so}(3)}(\bi{v})\, \circ = \dbracket{\,\circ \, }{\bi{v}}_{\mathfrak{so}(3)}^*
= \,\circ\, \times \bi{v} .
\label{Poisson_oprator_IX_for_fields}
\end{equation}

\subsection{$\mathrm{curl}$ deformation}
\label{subsec:curl}
We have yet to \emph{deform} the bracket $\dpoisson{~}{~}_{\mathfrak{so}(3)}$ in order to obtain the vortex dynamics bracket $\dpoisson{~}{~}_{\bomega} $\,\cite{YM2020}.
Remember the discussion in Sec.\,\ref{subsec:varity_3D};
changing the observable ($\in\mathfrak{g}^*$) induces a \emph{deformation} of the Lie algebra.
Here we observe $\bomega=\mathrm{curl}\,\bi{v}$ instead of $\bi{v}$,
i.e., 
we deform $\dangle{\dbracket{\partial_{\bi{v}} G}{\partial_{\bi{v}} H}_{\mathfrak{so}(3)}}{\bi{v}}$ to  $\dangle{\dbracket{\partial_{\bi{v}} G}{\partial_{\bi{v}}H }_{\mathfrak{so}(3)}}{\mathrm{curl}\, \bi{v} }$,
which means that we deform the Poisson operator
$\mathcal{J}_{\mathfrak{so}(3)}(\bi{v})\,\circ$ to
$\mathcal{J}_{\mathfrak{so}(3)}(\mathrm{curl}\,\bi{v})\,\circ$.

Since $\mathrm{curl}^{-1}$ is a surjection from $V$ to $V_{\bomega}$ of (\ref{domain_curl}), every $\bi{v}\in V_{\bomega}$ can be represented by $\bomega$\,\cite{compressible}.
% \footnote{
% This is not the case for compressible fluids (see Sec.\,\ref{sec:general_fluid}). }
So, we evaluate physical quantities as functionals of $\bomega$,
i.e., the phase space may be changed from $V$ to $V_{\bomega}$.
Then, by chain rule, we evaluate
\[
\partial_{\bi{v}} F (\bomega) = \mathrm{curl}\,\partial_{\bomega} F(\bomega).
\]
Using this, the curl-deformed $\dpoisson{~}{~}_{\mathfrak{so}(3)}$ reads
\begin{eqnarray}
\dpoisson{G}{H}_{\bomega} &:=& 
\dangle{\dbracket{\partial_{\bi{v}} G}{\partial_{\bi{v}} H}_{\mathfrak{so}(3)} }{\mathrm{curl}\,\bi{v} }
\nonumber \\
&=&
\dangle{ \dbracket{\mathrm{curl}\,\partial_{\bomega} G}{\mathrm{curl}\,\partial_{\bomega} H}_{\mathfrak{so}(3)} }{ \bomega }
\nonumber \\
&=&
\dangle{ \mathrm{curl}\, \partial_{\bomega} G }{\dbracket{\mathrm{curl}\,\partial_{\bomega} H}{\bomega}^*_{\mathfrak{so}(3)} }
\nonumber \\
&=&
\dangle{ \partial_{\bomega} G}{
\mathrm{curl}\, \dbracket{\mathrm{curl}\, \partial_{\bomega} H}{\bomega}^*_{\mathfrak{so}(3)} } .
\label{SO(3)_of_fields-curled-curled}
\end{eqnarray}
The corresponding Poisson operator is
\begin{equation}
\mathcal{J}_{\bomega}(\bomega)
= \mathrm{curl}\,[(\mathrm{curl}\,\,\circ\, )\times\bomega ],
% = \mathscr{P}_\Sigma \nabla\times( (\nabla\times\circ\,)\times\bomega),
\label{Poisson-operator_for_SO(3)_of_fields_curled-curled}
\end{equation}
which is the hoped-for vortex dynamics Poisson operator (\ref{vortex_Poisson_naive}).

%%%%%%%%%%

\subsection{Clebsch parameterization of the vorticity}
\label{subsec:Clebsch_vorticity}

In analogy of the angular-momentum reduction of $\mathfrak{sp}(6,\mathbb{R})$, and its reversed interpretation as the Clebsch parameterization of $\mathfrak{so}(3)$, 
we introduce the Clebsch parameterization of the vorticity $\bomega$ to formulate the canonical Poisson bracket that subsumes $\dpoisson{~}{~}_{\bomega}$ as the reduction\,\cite{MarsdenWeinstein1983}.

We consider the vector bundle of $\mathfrak{sp}(4,\mathbb{R})$ fibers over the base space $\Omega$, and write the canonical variables as
\[
\bzeta=({\alpha}_1(\bi{x}), {\alpha}_2(\bi{x}), {\beta}_1(\bi{x}), {\beta}_2(\bi{x})) \in V_{\bzeta}.
\]
Endowing with the canonical Poisson bracket 
\[
\dpoisson{G(\bzeta)}{H(\bzeta)}_{\mathfrak{sp}(4)} := \dangle{\partial_{\bzeta}G}{J_c \partial_{\bzeta}H },
\quad
J_c = \left( \begin{array}{cc}
0 & I \\
-I & 0
\end{array}\right),
\]
the function space $ V_{\bzeta}$ is a canonical Poisson manifold. 
We put
\begin{equation}
\bi{v} = {\alpha}_1 \rmd {\beta}_1 + {\alpha}_2 \rmd {\beta}_2 ,
\label{Clebsch_for_omega}
\end{equation}
and use this as the Clebsch parameterization of the fluid velocity (in fact, the momentum 1-form).
The vorticity $\bomega=\nabla\times\bi{v}$ is parameterized as the 2-form (see Remark\,\ref{remark:duality})
\begin{equation}
\bomega = \rmd {\alpha}_1 \wedge \rmd {\beta}_1 + \rmd {\alpha}_2 \wedge \rmd {\beta}_2 .
\label{Clebsch_for_omega-2}
\end{equation}
See (\ref{angular_momentum}) for the analogy of the angular-momentum parameterization.
By
\[
\tilde{\bomega} = \sum_j \rmd \tilde{\alpha}_j \wedge \rmd {\beta}_j+ \rmd {\alpha}_j \wedge \rmd \tilde{\beta}_j 
=  \sum_j \rmd (  \tilde{\alpha}_j \rmd {\beta}_j - \tilde{\beta}_j \rmd {\alpha}_j ),
\]
we observe
\begin{eqnarray*}
\tilde{F} &=& \sum_j  \dangle{\partial_{{\beta}_j} F}{\tilde{{\beta}_j}} + \dangle{\partial_{{\alpha}_j}F}{\tilde{{\alpha}_j}}
\\
&=& \dangle{\partial_{\bomega} F}{\tilde{\bomega}}
\\
&=& \sum_j \dangle{\partial_{\bomega} F}{~\rmd ( \tilde{\alpha}_j \rmd {\beta}_j - \tilde{\beta}_j \rmd {\alpha}_j  )}
\\
&=&  \sum_j \dangle{\delta \partial_{\bomega} F}{~\tilde{\alpha}_j  \rmd {\beta}_j - \tilde{\beta}_j \rmd {\alpha}_j } ,
\end{eqnarray*}
where $\delta = * \rmd * $ formally applies as $\nabla\times$, and maps 2-forms to 1-forms.
Therefore, we obtain
\begin{equation}
\left\{ \begin{array}{l}
\partial_{{\beta}_j} F = - \delta \partial_{\bomega} F \wedge \rmd {\alpha}_j ,
\\ 
\partial_{{\alpha}_j} F = ~~ \delta \partial_{\bomega} F \wedge \rmd {\beta}_j .
\end{array} \right.
\label{chain_omega}
\end{equation}
Using this (and the differential geometrical formulation explained in Remark\,\ref{remark:duality}),
the canonical bracket $\dpoisson{G}{H}_{\mathfrak{sp}(4)}$ evaluates
for the Clebsch parameterized $G(\bomega)$ and $H(\bomega)$ as
\begin{eqnarray*}
\dpoisson{G(\bomega)}{H(\bomega)}_{\mathfrak{sp}(4)} &=& 
\sum_j  \int_\Omega \partial_{{\beta}_j}G \wedge * \partial_{{\alpha}_j} H - \partial_{{\alpha}_j}G \wedge * \partial_{{\beta}_j} H
\\
&=& 
\sum_j  \int_\Omega - (\delta \partial_{\bomega} G \wedge \rmd {\alpha}_j  )\wedge * (\delta \partial_{\bomega} H \wedge \rmd {\beta}_j )
\\
&~& ~~~~~~~~
+  (\delta \partial_{\bomega} G \wedge \rmd {\beta}_j  )\wedge * (\delta \partial_{\bomega} H \wedge \rmd {\alpha}_j )
\\
&=& 
\dangle{ \dbracket{\delta\partial_{\bomega} G}{\delta\partial_{\bomega} H}_{\mathfrak{so}(3)} }{ \bomega }
\\
&=& \dpoisson{G}{H}_{\bomega}
\end{eqnarray*}
Therefore, the vortex dynamics bracket $\dpoisson{~}{~}_{\bomega}$ is the reduction of the canonical bracket $\dpoisson{~}{~}_{\mathfrak{sp}(4)}$ for the vorticity $\bomega$ defined as (\ref{Clebsch_for_omega-2}).
Conversely, 
the noncanonical bracket $\dpoisson{~}{~}_{\bomega}$ is canonicalized as $\dpoisson{~}{~}_{\mathfrak{sp}(4)}$ by the Clebsch parameterization (\ref{Clebsch_for_omega-2}).

%%%%%%%%%%%%%%%%%%%%%%%%%%%%%%%%%%%%%%%%%%%%%%%%%%%%%%%%%%%%%%%%%%%%%%%%%%%%%%%%%%%%%%%%%%%%%%%%%%%%%%%%%%%%%%%
%%%%%%%%%%%%%%%%%%%%%%%%%%%%%%%%%%%%%%%%%%%%%%%%%%%%%%%%%%%%%%%%%%%%%%%%%%%%%%%%%%%%%%%%%%%%%%%%%%%%%%%%%%%%%%%
%%%%%%%%%%%%%%%%%%%%%%%%%%%%%%%%%%%%%%%%%%%%%%%%%%%%%%%%%%%%%%%%%%%%%%%%%%%%%%%%%%%%%%%%%%%%%%%%%%%%%%%%%%%%%%%
\section{General fluid dynamics}
\label{sec:general_fluid}
Although the vortex equation of incompressible fluid is a good example of Nambu dynamics,
it falls short of discussing the quantization.
We  prepare a more general compressible fluid model, and show that the fluid's noncanonical Poisson algebra is canonicalized by a more general Clebsch parameterization that can be naturally quantized.
The generalized formulation goes almost parallel to that of the preceding section.

\subsection{Basic equation}
\label{subsec:compressible_fluid}
The domain of the fluid is the same $\Omega=\mathbb{T}^3$.
We consider an ideal (inviscid and barotropic) fluid, and
denote the mass density by $\rho$ and the fluid velocity by $\bi{v}$.
Instead of the constant density and the incompressibility condition $\nabla\cdot\bi{v}=0$ in the vortex dynamics model of Sec.\,\ref{subsec:vortex_equation},
we invoke the mass conservation law
\begin{equation}
\partial_t{\rho} = -\nabla\cdot(\bi{v} {\rho}).
\label{mass_conservation}
\end{equation}

The fluid velocity is governed by the momentum equation is
\begin{equation}
\partial_t \bi{v}=-(\bi{v}\cdot\nabla)\bi{v} - \rho^{-1}\nabla P,
\label{momentum}
\end{equation}
where $P$ is the pressure that is a function of $\rho$ (barotropic relation),
hence we may write $\rho^{-1}\nabla P = \nabla h(\rho)$ with a scalar function $h(\rho)$ that corresponds to the enthalpy of the fluid.

\subsection{Hamiltonian form of compressible fluid mechanics}
\label{subsec:compressible_fluid_Hamiltonian}
We can cast the ideal fluid equations (\ref{mass_conservation})-(\ref{momentum}) into a Hamiltonian form.
The \emph{fluid variables} are denoted by
\[
\bi{u} = (\rho, \bi{v})^{\textrm{T}} \in \mathcal{U},
\]
and $\mathcal{U}$ will be called the fluid phase space, which is made to be $L^2(\Omega)$ with the standard inner product $\dangle{\bi{a}}{\bi{b}} = \int_\Omega \bi{a}\cdot\bi{b}\,\rmd^3 x$.
We consider the Poisson operator
\begin{equation}
 J_F(\bi{u}) = 
\left( \begin{array}{cc} 0 & -\nabla\cdot \\
-\nabla & -\left(\frac{\nabla\times\bi{v}}{\rho}\right) \times
 \end{array} \right)  ,
\label{fluid_Poisson}
\end{equation}
by which we define the ``fluid Poisson bracket''
\begin{equation}
\dpoisson{G(\bi{u})}{H(\bi{u})}_F 
=  \dangle{\partial_{\bi{u}} G}{ J_F (\bi{u}) \partial_{\bi{u}} H }.
\label{bracket-F_derivation}
\end{equation}
Since $J_F(\bi{u})$ is not a linear function of $\bi{u}$, it is not the type of Lie-Poisson.  
However, it is known that $\dpoisson{~}{~}_F $ satisfies the requirements of Poisson bracket\,\cite{Morrison1998,Morrison1982}.
Evidently, the total mass $N=\int_{\Omega} \rho\,\rmd^3x$ and the helicity
\begin{equation}
C = \int_{\Omega} \bi{v}\cdot\nabla\times\bi{v}\,\rmd^3 x 
\label{Helicity_naive}
\end{equation}
are the Casimirs of $\dpoisson{~}{~}_F $.
Here $C$ is evaluated as a functional of $\bi{v}$; compare with (\ref{helicity_of_vorticity}).

The fluid energy defines the Hamiltonian:
\begin{equation}
H(\bi{u}) = \int_{\Omega} \left[ \frac{1}{2} |\bi{v}|^2 + \varepsilon(\rho)  \right]\,\rho\,\rmd^3 x,
\label{fluid_energy}
\end{equation}
where $\varepsilon(\rho)$ is the specific internal (thermal) energy, satisfying $\partial(\rho\varepsilon(\rho))/\partial\rho = h(\rho)$ (specific enthalpy;
assuming  a barotropic pressure $P=P(\rho)$, we have $ \rho^{-1}\nabla P = \nabla h$).
We easily find that Hamilton's equation, $\partial_t \bi{u} = J_F(\bi{u}) \partial_{\bi{u}} H (\bi{u})$, with the Hamiltonian (\ref{fluid_energy}) reproduces the fluid equations (\ref{mass_conservation})-(\ref{momentum}).

\begin{remark}[isentropic fluid]
\label{remark:entropy}
\normalfont
Since this paper is aimed at formulating a fluid mechanical Nambu dynamics, we assume the simple ``barotropic'' relation in the Hamiltonian (\ref{fluid_energy}) so that the helicity (\ref{Helicity_naive}) is conserved as the Casimir.
In a more general setting of ideal fluid mechanics, we may consider the isentropic fluid\,\cite{Morrison1998}.
Including the specific entropy $\sigma$ as an additional state variable, we consider the extended fluid variables $\tilde{\bi{u}}=(\rho,\bi{v},\sigma)^{\mathrm{T}}$, and the internal energy such as $\varepsilon(\rho,\sigma)$.
With the thermodynamic temperature $T=\partial\varepsilon/\partial\sigma$, we may put $\rho\nabla P = \nabla h - T \nabla\sigma$.
The Poisson operator (\ref{fluid_Poisson}) is extended to 
\begin{equation}
 \tilde{J}_{F}(\tilde{\bi{u}}) = 
\left( \begin{array}{ccc} 0 & -\nabla\cdot & 0\\
-\nabla & -\left(\frac{\nabla\times\bi{v}}{\rho}\right) \times & \frac{\nabla\sigma}{\rho} \\
0 & - \frac{\nabla\sigma}{\rho} \cdot & 0
 \end{array} \right)  .
\label{fluid_Poisson-ex}
\end{equation}
The last row of $ \tilde{J}_{F}$ describes the entropy conservation $\partial_t\sigma=-\bi{v}\cdot\nabla\sigma$.
\end{remark}

\subsection{Vector bundle of $\mathfrak{sp}(6,\mathbb{R})$ fibers}
\label{subsec:compressible_fluid_canonical_variables}
Now we embed the fluid phase space $\mathcal{U}$, the Poisson manifold endowed with the noncanonical bracket $\dpoisson{~}{~}_F $, into a larger Poisson manifold for canonicalization.

Over the base space ${\Omega}=\mathbb{T}^3$, we define the vector bundle $E_{\Omega\times\mathfrak{sp}(6,\mathbb{R})}$ of the $\mathfrak{sp}(6,\mathbb{R})$ fibers. 
We represent the canonical variables by
\begin{equation}
% \bxi  =  (\varphi, q,s,\varrho,  p,  r)^{\mathrm{T}} ~\in X,
% \bxi  =  (\varphi, q,s,\varrho,  p,  r)^{\mathrm{T}} ~\in X,
% \bxi  =  (\varphi, \varrho, q_1, p_1, q_2, p_2)^{\mathrm{T}} ~\in X,
% \bxi  =  (\varphi, q_1, q_2, \varrho, p_1, p_2)^{\mathrm{T}} ~\in X,
% \bxi  =  ( \varrho, p_1, p_2, \varphi, q_1, q_2)^{\mathrm{T}} ~\in X,
\bxi  =  ( \varrho, {\alpha}_1, {\alpha}_2, \varphi, {\beta}_1, {\beta}_2)^{\mathrm{T}} ~\in X,
\label{canonical_variables}
\end{equation}
where $\xi_1=\varrho, \xi_2={\alpha}_1, \xi_3={\alpha}_2$ are 3-forms, and $\xi_4=\varphi, \xi_5={\beta}_1, \xi_6 ={\beta}_2$ are 0-forms (scalars) in the base space ${\Omega}$.
The dual space $X^*$ is the Hodge-dual of $X$, i.e., the first three components of $\beta\in X^*$ are 0-forms and the remaining three components are 3-forms.
The pairing of $X^*$ and $X$ is\,\cite{Clebsch_forms}
% \footnote{
% Here the components of $\bxi \in X$ are mixture of 0-forms and 3-forms (for the convenience of the latter formulation).  So the duality is defined in a different way in comparison with the definition in Remark\,\ref{remark:duality}.
%  }
\begin{equation}
\dangle{\beta}{\bxi }= \sum_{j} \int_{\Omega} \eta^j  \xi_j ,
\quad \beta\in X^*, ~ \bxi\in X.
\label{pairing}
\end{equation}
The gradient of $F(\bxi)$ is defined by $\tilde{F}= \dangle{\partial_{\bxi} F}{ \tilde{\bxi} } $,
so $\partial_{\bxi} F \in X^*$.
For $G,~H\in C^\infty(X)$, we define a canonical Poisson bracket
\begin{equation}
\dpoisson{G}{H}_{\mathfrak{sp}(6)} = \dangle{\partial_{\bxi} G}{J_c \partial_{\bxi} H } ,
\label{canonical_Poisson_bracket}
\end{equation}
where $J_c :\, X^* \rightarrow X$ is the symplectic operator 
\begin{equation}
J_c = \left( \begin{array}{cc}
0 & I \\
-I & 0
\end{array}\right).
\label{symplectic}
\end{equation}

\subsection{Clebsch parameterization of the fluid variables}
\label{subsec:compressible_fluid_Clebsch}

We relate the fluid variable $\bi{u}\in \mathcal{U}$ and the abstract field $\bxi\in X$ by
\begin{eqnarray}
\rho &=& \varrho^* ,
\label{Clebsch0}
\\
\bi{v} &=& \rmd\varphi + \check{{\alpha}_1} \rmd {\beta}_1 + \check{{\alpha}_2} \rmd {\beta}_2 ,
\quad \left(\check{{\alpha}_1}=\frac{{\alpha}_1}{\varrho}, ~\check{{\alpha}_2}=\frac{{\alpha}_2}{\varrho} \right) ,
\label{Clebsch}
\end{eqnarray}
where $\varrho^*$ is the Hodge dual (scalar part) of the 3-form $\varrho$.
% \begin{equation}
% \rho \Leftrightarrow \varrho^* ,
% \label{Clebsch0}
% \end{equation}
% where we denote by $\varrho^*$ the Hodge dual (scalar part) of the 3-form $\varrho$,
% (i.e. $\varrho^* \vol^3 = \varrho$),
% and  
% \begin{equation}
% \bi{V} \Leftrightarrow    \wp = \rmd\varphi + \check{p} \rmd q + \check{r} \rmd s ,
% \quad \left(\check{p}={p^*}/{\varrho^*}, ~\check{r}={r^*}/{\varrho^*} \right).
% \label{Clebsch}
% \end{equation}
Writing the 1-form $\bi{v}$ as (\ref{Clebsch}) is called the \emph{Clebsch parameterization}\,\cite{footnote:Clebsch}.
% \footnote{\label{footnote:Clebsch}
% The original idea of Clebsch\,\cite{Clebsch1859} was to modify a curl-free vector field $\bi{a}=\nabla \varphi$ to $\bi{b}=\nabla \varphi + {\alpha}_1\nabla {\beta}_1$ to yield a finite curl $\nabla \alpha_1 \times \nabla \beta_1$.  
% However, an arbitrary 3-vector may not be written in the form of $\bi{b}$; we have to add another non-exact term ${\alpha}_2\nabla {\beta}_2$ to parameterize every 3-vector\,\cite{Clebsch}.
% }

For $F(\bi{u}(\bxi))$, we may evaluate (writing in vector analysis notation)
\[
\left\{ \begin{array}{l} \partial_\varrho F = \partial_\rho F - \frac{1}{\rho}  (\bi{v}-\nabla \varphi) \cdot \partial_{\bi{v}} F ,
\\
\partial_\varphi F = -\nabla\cdot(\partial_{\bi{v}} F) ,
\end{array} \right.
\quad
\left\{ \begin{array}{l} 
\partial_{\alpha_j} F = \frac{1}{\rho} \nabla\beta_j \cdot(\partial_{\bi{v}} F) ,
\\
\partial_{\beta_j} F = -\nabla\cdot \left( \frac{\alpha_j}{\rho} \partial_{\bi{v}} F \right).
\end{array} \right.
\]
Using these relations, we find, for $G(\bi{u}(\bxi))$ and $H(\bi{u}(\bxi))$,
\begin{equation}
\dpoisson{G(\bi{u}(\bxi))}{H(\bi{u}(\bxi))}_{\mathfrak{sp}(6)} = \dpoisson{G(\bi{u})}{H(\bi{u})}_F.
\label{fluid_reduction_of_sp(6)}
\end{equation}

Let us represent the fluid equations (\ref{mass_conservation})-(\ref{momentum}) in terms of the Clebsch parameters.
Inserting (\ref{Clebsch0})-(\ref{Clebsch}) into the fluid energy (\ref{fluid_energy}), we obtain
\begin{equation}
H(\bi{u}(\bxi)) = \int_{\Omega} \left[  \frac12 { \big| \rmd\varphi + \check{\alpha}_1\rmd {\beta}_1 + \check{\alpha}_2\rmd {\beta}_2 \big|^2}
+ \varepsilon(\varrho^*)  \right]\,\varrho \,.
\label{Hamiltonian}
\end{equation}
Using this $H(\bxi)$ as the Hamiltonian\,\cite{variational_principle},
% \footnote{
% The action principle for the fluid dynamics equations was first studied by Lin\,\cite{Lin}, 
% and followed by other authors \cite{Zakharov,Seliger,Salmon}
% (see also \cite{Jackiw,Jackiw2} for the discussion in the context of particle physics).
% In order to produce a finite vorticity, one has to introduce some Lagrange multipliers for (abstract) constraints on the streamlines, and these multipliers alchemize into the Clebsch parameters.
% }
Hamilton's equation $\partial_t \bxi = J_c \partial_{\xi} H(\bxi)$ reads
% (denoting by $L_{\bi{V}}$ the Lie derivative along the vector $\bi{V}\in T{M}$)
\begin{eqnarray}
& & (\partial_t + \mathcal{L}_{\bi{v}}) \varrho = 0, \quad (\partial_t + \mathcal{L}_{\bi{v}}) {\alpha}_1 = 0, \quad (\partial_t + \mathcal{L}_{\bi{v}}) {\alpha}_2 = 0 , 
\label{H-3} 
\\
& & (\partial_t + \mathcal{L}_{\bi{v}})\varphi = \frac{1}{2} {v^2} - h,
\quad (\partial_t + \mathcal{L}_{\bi{v}}) {\beta}_1 = 0, \quad (\partial_t + \mathcal{L}_{\bi{v}}) {\beta}_2 = 0,
\label{H-2} 
\end{eqnarray}
where ${\mathcal{L}}_{\bi{v}}$ is the Lie derivative (by the contravariant counterpart of the covariant $\bi{v}$).
% where $\widetilde{\mathcal{L}}_{\bi{V}}$ is the (non-relativistic) space-time Lie derivative:
% \begin{equation}
% \widetilde{\mathcal{L}}_{\bi{v}} = \partial_t + \mathcal{L}_{\bi{v}},
% \label{extended_Lie_detivative}
% \end{equation}
% with $\mathcal{L}_{\bi{V}}$ being  the conventional Lie derivative along the vector $\bi{V}\in T{\Omega}$.
% \footnote{
% Here $\bi{V}$ is regarded as a vector $\in T\Omega$ through the following identification.
% By (\ref{Clebsch}), $\bi{V} \Leftrightarrow \wp \in T^*\Omega$.
% In the Hamiltonian (\ref{Hamiltonian}), $|\wp|^2 \Leftrightarrow \bi{V}^\dagger\cdot\bi{V}$ with the dual
% $\bi{V}^\dagger=\bi{V}\in T\Omega$.
% We point out that a similar Lie derivative construction exists for extended MHD \cite{LMM16} in its natural (physical) variables.
% }
The first equation of (\ref{H-3}) is nothing but the mass conservation law (\ref{mass_conservation}).
Evaluating $\partial_t\bi{v}$ by inserting (\ref{Clebsch}) and using (\ref{H-2})-(\ref{H-3}), we
obtain the momentum equation of ideal fluid (\ref{momentum}).

In summary, the canonical Poisson bracket $\dpoisson{~}{~}_{\mathfrak{sp}(6)}$, when evaluated for $H(\bi{u})$, reduces to the fluid Poisson bracket  $\dpoisson{~}{~}_F$.
Or, conversely, the noncanonical fluid bracket $\dpoisson{~}{~}_F$ is canonicalized as $\dpoisson{~}{~}_{\mathfrak{sp}(6)}$ by the Clebsch parameterization  (\ref{Clebsch0})-(\ref{Clebsch}).

\subsection{Sub-algebras of the general fluid system}
\label{subsec:subsystems}
A merit of formulating the canonical Hamiltonian system (and the underlying canonical Poisson algebra) is in that we may delineate a variety of closed ``subsystems'' by symmetries in the phase space.

\begin{enumerate}
\item
\textbf{Incompressible flow}:
When we suppress the variable $\varphi$ in the phase space, the flow velocity of (\ref{Clebsch}) reduces to
$\bi{v}= \check{{\alpha}_1} \rmd {\beta}_1 + \check{{\alpha}_2} \rmd {\beta}_2 $.
By $\partial_\varphi H = 0$, the conjugate variable $\rho$ becomes constant,
and hence, the above $\bi{v}$ is equivalent to (\ref{Clebsch_for_omega}) of the vortex dynamics.
The remaining part of the fluid equation $\partial_t \bi{u} = J_F \partial_{\bi{u}} H$ reads (putting $\rho=1$)
\[
\partial_t \bi{v} = -( \nabla\times\bi{v}) \times \bi{v} ,
\]
which is equivalent to the vortex dynamics equation (\ref{vortex_eq}) through the relation $\bi{v} = \mathrm{curl}^{-1} \bomega$.
The Clebsch parameters obey remarkably simple equations:
\[
(\partial_t + \mathcal{L}_{\bi{v}}) {\alpha}_j = 0, ~~
%  (\partial_t + \mathcal{L}_{\bi{v}}) {\alpha}_2 = 0 , 
 (\partial_t + \mathcal{L}_{\bi{v}}) {\beta}_j = 0, 
% ~~ (\partial_t + \mathcal{L}_{\bi{v}}) {\beta}_2 = 0,
\quad (j=1,2)
\]
implying that all fields are simply Lie-dragged
(remember (\ref{H-3})-(\ref{H-2}), where the nontrivial equation only pertains to $\varphi$).

\medskip
\item
\textbf{Epi-2D flow}:
Removing $\alpha_2$ (or $\beta_2$), the flow velocity is reduced to the form such that
\begin{equation}
\bi{v}= \rmd\varphi + \alpha_1 \rmd \beta_1 
= \nabla \varphi + \alpha_1 \nabla \beta_1,
\label{epi-2D}
\end{equation}
which we call \emph{epi-2 dimensional} flow\,\cite{YoshidaMorrison_epi2D}.
As remarked in footnote\,\cite{footnote:Clebsch}, epi-2 dimensional 1-forms constitutes a special class of 1-forms in 3-dimensional space.
Especially, the corresponding vorticity $\bomega= \rmd \alpha_1 \wedge \rmd \beta_1 = \nabla \alpha_1 \times \nabla \beta_1$ has only integrable vortex lines; the level sets of both $\alpha_1(\bi{x})$ and $\beta_1(\bi{x})$ are the integral surfaces.

\medskip
\item
\textbf{Irrotational flow}: 
When $\alpha_1$ (or $\beta_1$) also is removed, the flow velocity (in fact, the momentum 1-form) becomes an exact 1-form
\begin{equation}
\bi{v}= \rmd \varphi = \nabla \varphi ,
\label{HJ-1}
\end{equation}
and then, $\bi{v} \in \mathrm{Ker(curl)}$. 
We also separate the harmonic fields (such that $\nabla\times\bi{v}=0$ and $\nabla\cdot\bi{v}=0$) from the phase space as well as the state space.
The Poisson operator (\ref{fluid_Poisson}) reduces to
\begin{equation}
 J_{icF}(\bi{u}) = 
\left( \begin{array}{cc} 0 & -\nabla\cdot \\
-\nabla &0
 \end{array} \right)  ,
\label{fluid_Poisson-ic}
\end{equation}
which only has a unique Casimir $N=\int_\Omega \rho\,\rmd^3 x$ (total mass).
In the Clebsch parameterized equation of motion (\ref{H-3}) and (\ref{H-2}), the first equations remain nontrivial;
(\ref{H-3}) is the mass conservation law, and (\ref{H-2}) reads
\begin{equation}
\partial_t \varphi = -\left( \frac{1}{2} |\bi{v}|^2 + h \right).
\label{HJ-0}
\end{equation}

\end{enumerate}

By interpreting $\varphi=$ action, and $\frac{1}{2} |\bi{v}|^2 + h =$ Hamiltonian, we may regard (\ref{HJ-1})-(\ref{HJ-0}) as the Hamilton-Jacobi equations.
In the next section, this relation will play the role of  bridging fluid mechanics and quantum field theory.
And the generalized form of the momentum such as (\ref{epi-2D}) and (\ref{Clebsch}), which involves \emph{vorticity}, will introduce an interesting structure.

%%%%%%%%%%%%%%%%%%%%%%%%%%%%%%%%%%%%%%%%%%%%%%%%%%%%%%%%%%%%%%%%%%%%%%%%%%%%%%%%%%%%%%%%%%%%%%%%%%%%%%%%%%%%%%%%%%%%
%%%%%%%%%%%%%%%%%%%%%%%%%%%%%%%%%%%%%%%%%%%%%%%%%%%%%%%%%%%%%%%%%%%%%%%%%%%%%%%%%%%%%%%%%%%%%%%%%%%%%%%%%%%%%%%%%%%%
%%%%%%%%%%%%%%%%%%%%%%%%%%%%%%%%%%%%%%%%%%%%%%%%%%%%%%%%%%%%%%%%%%%%%%%%%%%%%%%%%%%%%%%%%%%%%%%%%%%%%%%%%%%%%%%%%%%%
\section{Quantization through Madelung representation}
\label{sec:Madelung}
Elucidating the relation between the \emph{Madelung representation} of matter waves and the Clebsch parameterization,
we discuss the quantization of Nambu mechanics.

\subsection{Eikonal of the wave function and the Hamilton-Jacobi equation}
Madelung's original idea was to represent the matter wave function $\psi$ by fluid-like parameters $(\rho, \bi{v})$ and describe quantum mechanics in the analogy of fluid mechanics\,\cite{Madelung}.
By putting
\begin{equation}
\psi (\bi{x},t) = \sqrt{\rho(\bi{x},t)} e^{i\varphi(\bi{x},t)/\hbar} ,
\label{Madelung-1}
\end{equation}
the real and imaginary parts of the complex function $\psi$ are mapped to the two real functions, the \emph{density} $\rho= \psi^*\psi$ and  the \emph{eikonal} (or \emph{action}) $\varphi$.
We define the momentum (normalizing the mass to 1, we write the momentum $\bi{p}$ as the velocity $\bi{v}$)
\begin{equation}
\bi{v} =\Re \frac{\psi^* \cdot (-i\hbar \nabla\psi) }{\rho} = \nabla \varphi ,
% \bi{v} = \frac{ -i\hbar \nabla\psi\cdot\psi^* }{\rho} = \nabla \varphi ,
\label{Madelung-1-2}
\end{equation}
which parallels the quantization rule $\bi{p} \Rightarrow -i\hbar\partial_{\bi{x}}$.

Let us see how the \emph{correspondence principle} works.
For a quantum mechanical quantity
$\mathcal{G}$, which is a self-adjoint operator acting on the Hilbert space $L^2(\Omega)$,
its measurement is evaluated by
\[
G = \dangle{\psi}{\mathcal{G} \psi} = \int_\Omega \psi^* \cdot (\mathcal{G} \psi )\,\rmd^3 x.
\]

Given a Hamiltonian $\mathcal{H}$,
the wave function obeys the Schr\"odinger equation
\begin{equation}
i\hbar \partial_t \psi = \mathcal{H} \psi ,
\label{Schoedinger}
\end{equation}
which is formally solved as $\psi(t) = \rme^{t\mathcal{H}/i\hbar}\psi_0$.
For a time-independent observable $\mathcal{G}_0$, we define
$ G(t)=\dangle{\psi(t)}{\mathcal{G}_0 \psi(t)} = \dangle{\psi_0}{\mathcal{G}(t) \psi_0}$,
i.e., 
$\mathcal{G}(t) = \rme^{-t\mathcal{H}/i\hbar} \mathcal{G}_0 \rme^{t\mathcal{H}/i\hbar}$,
% \footnote{
% In the following discussion, we will encounter nonlinear generator $\mathcal{H}$, for which we may not assume 
% $\mathcal{H}\rme^{t\mathcal{H}/i\hbar} = \rme^{t\mathcal{H}/i\hbar} \mathcal{H}$.
% Then, in (\ref{Heisenberg_eq-0}) and (\ref{Heisenberg_eq}), we evaluate the time derivatives at $t=0$. 
% }
% Then, 
% \begin{equation}
% \dot{G} = \dangle{\dot{\psi}}{\mathcal{G}_0 \psi} + \dangle{\psi}{\mathcal{G}_0\dot{\psi}} = 
% \dangle{\psi_0}{\dot{\mathcal{G}} \psi_0} ,
% \label{Heisenberg_eq-0}
% \end{equation}
% which implies the Heisenberg equation
which satisfies the Heisenberg equation\,\cite{generator}
% \footnote{
% The temporal derivative $\dot{\mathcal{G}}$ (or $\dot{G}$) is evaluated at $t=0$.
% }
\begin{equation}
\dot{\mathcal{G}} = \frac{1}{i\hbar} [\mathcal{G}, \mathcal{H} ].
\label{Heisenberg_eq}
\end{equation}
The correspondence principle demands that (\ref{Heisenberg_eq}) is equivalent to the classical equation $\dot{G} = \dpoisson{G}{H}$ in the limit $\hbar\rightarrow0$.

The classical Hamiltonian (\ref{fluid_energy}) is quantized by $\bi{p} \Rightarrow -i\hbar\partial_{\bi{x}}$:
\begin{eqnarray}
{H}_{\mathrm{q}} 
% =  \dangle{\psi}{\mathcal{H} \psi} =
&=& \int_{\Omega} \psi^* \left( -\frac{\hbar^2}{2} \nabla^2 +  \varepsilon \right) \psi\,\rmd^3 x
\nonumber \\
&=& \int_{\Omega} \left[ \frac{1}{2} |\nabla\varphi|^2 + \varepsilon + \hbar^2 \frac{|\nabla\rho|^2}{8\rho^2}  \right]\,\rho\,\rmd^3 x,
\label{fluid_energy-q}
\end{eqnarray}
The action is $\int (\Theta - H_{\textrm{q}})\,\rmd t$ with
\begin{equation}
% \frac{1}{2} 
\Theta = \Re \dangle{\psi}{i\hbar \partial_t\psi} = -\int \rho(\partial_t\varphi)\,\rmd^3x .
\label{Q-Hamiltonian-Madelung-2}
\end{equation}
When represented by $\psi$, the variational principle yields the Schr\"odinger equation (\ref{Schoedinger})
with the quantized Hamiltonian\,\cite{nonlinearity}
% \footnote{
% Notice that $H_{\mathrm{q}}$ of (\ref{fluid_energy-q}) is not $\dangle{\psi}{\mathcal{H}\psi}$, reflecting the nonlinearity due to $\varepsilon(|\psi|^2)$.
% }
\begin{equation}
\mathcal{H} = 
- \frac{\hbar^2}{2} \nabla^2 + h ,
\label{Hamilton_operator}
\end{equation}
while, by $(\rho,\varphi)$, we obtain
\begin{eqnarray}
\partial_t \rho &=& - \nabla\cdot(\bi{v} \rho),
\label{Qfluid-1} \\
\partial_t \varphi &=& - \left( \frac{1}{2}|\bi{v}|^2 +h
- \frac{\hbar^2}{2} \frac{\nabla^2\sqrt{\rho}}{\sqrt{\rho}}\right).
\label{Qfluid-2} 
\end{eqnarray}
On the right-had side of (\ref{Qfluid-2}), the term multiplied by $\hbar^2$ reflects the quantum-mechanical effect,
which is called a \emph{quantum pressure}.
In the limit $\hbar\rightarrow0$, (\ref{Qfluid-2}) reduces to (\ref{HJ-0}), and then,
the \emph{Clebsch parameters} $(\rho,\varphi)$ obey the Poisson algebra of the irrotational flow (Sec.\,\ref{subsec:subsystems}-(3)).
To put it in another way, the system (\ref{Qfluid-1})-(\ref{Qfluid-2}) is the quantization of the classical irrotational fluid system.

The irrotational momentum $\bi{v}=\nabla\varphi$ is due to the fundamental fact that the wave function $\psi$ has a single valued phase $\varphi$.
For a quantum mechanical system to have a finite vorticity, therefore, $\psi$ must have phase singularities,
i.e., $\psi=0$ occurs at some points, so that the phase $\varphi$ can include angular components (cohomologies) circulating the zero points of $\psi$.
This gives ``quantized'' (or discrete) vorticity.
The other possibility of quantum vorticity, as shown in the next subsection, emerges in the multi-component (spinor) wave functions, which correspond to the epi-2 dimensional and 3 dimensional general (or incompressible) fluid models given in Sec.\,\ref{subsec:subsystems}.

\begin{remark}[nonlinear Schr\"odinger equation]
\label{remark:entropy-2}
\normalfont
In the Hamiltonian (\ref{Hamilton_operator}), we assume a nonlinear potential energy $\varepsilon(\rho)$ (representing the internal energy), so that the measurement of the energy becomes consistent to the fluid energy (\ref{fluid_energy}).
Notice that the Schr\"odinger equation (\ref{Schoedinger}) is nonlinear.
The Gross-Pitaevskii equation is an example of such a nonlinear Schr\"odinger equation, where $\psi$ represents the order parameter of the Bose-Einstein condensate. 
If the potential energy $\varepsilon$ is simply a function of $\bi{x}$ ($\in\Omega$), the Hamilton's operator is linear, and $h$ in (\ref{Hamilton_operator}) and (\ref{Qfluid-2}) can be replaced by $\varepsilon(\bi{x})$.
\end{remark}

\subsection{Vorticity of spinor wave functions}
To formulate finite vorticity quantum fields (that are the quantization of the fluid Poisson algebras), 
we start with the two-component spinor field $\Psi=(\psi_1, \psi_2)$
(the idea was proposed in the pioneering work of Takabayasi\,\cite{tak1,tak2,tak3}).
The generalized Madelung representation of the wave function
\begin{equation}
\psi_j (\bi{x},t) = \sqrt{\rho_j(\bi{x},t)} \rme^{i\mathscr{S}_j(\bi{x},t)/\hbar}
\quad (j=1,2) .
\label{Madelung-2}
\end{equation}
converts the two complex field variables $(\psi_1, \psi_2)^{\mathrm{T}}$ into the four real variables $(\rho_1,\mathscr{S}_1,\rho_2,\mathscr{S}_2)$.
By linear transformation, we convert the Madelung parameters to a set of Clebsch parameters:
\begin{equation}
\left\{ \begin{array}{l}
\rho = \rho_1+\rho_2, \\%= \Psi^*\Psi\\,
{\alpha}= \rho_1-\rho_2, 
\end{array} \right.
\quad
\left\{ \begin{array}{l}
\varphi = \frac{1}{2} (\mathscr{S}_1+\mathscr{S}_2), \\
{\beta} = \frac{1}{2} (\mathscr{S}_1-\mathscr{S}_2). 
\end{array} \right.
\label{Clebsch-1}
\end{equation}
The momentum is
\begin{equation}
\bi{v}  = \Re \frac{\Psi^* \cdot (-i\hbar \nabla\Psi) }{\rho} = 
\nabla\varphi +  \frac{{\alpha}}{\rho} \nabla {\beta} ,
\label{Clebsch2}
\end{equation}
which produces a finite vorticity:
$\bomega =\nabla\times\bi{v}  =\nabla (\alpha/\rho)\times \nabla {\beta}$.

% \textcolor{red}{
% \begin{remark}[singularity]
% In (\ref{spin_vorticity2''}), we assumed that $\varphi$ does not have a phase singularity.
% If $\varphi$ is a multi-valued function, having phase singularity along \emph{vortex filaments}, 
% a finite vorticity can be created without ${\alpha}$ and ${\beta}$,
% or a finite helicity can be created when the vortex filaments have links.
% It may be controvertible whether a scalar quantum field (or spin 0 field) can produce linked vortex filaments.
% \end{remark}
% }

% \subsection{Decomposition of Hamiltonian into classical and quantal parts}
Let us examine the correspondence principle.
To see how the Clebsch parameters relate the Lie algebra of quantized observables and the canonical Poisson algebra of classical fields,
a simple Hamiltonian suffices:
\begin{eqnarray}
H_{\mathrm{q}} 
% &=& \dangle{\Psi}{\mathcal{H}\Psi} \nonumber \\
&=&
\int \Psi^* \cdot \left[ \left(- \frac{\hbar^2}{2} \nabla^2 + \varepsilon \right) \Psi
\right]
\,\rmd^3 x 
\nonumber \\
&=& \int \left[ \frac{1}{2}|\nabla \varphi + \check{\alpha}\nabla{\beta}|^2
+ \varepsilon 
+ \frac{\hbar^2}{8} \left(  \frac{|\nabla\rho|^2}{\rho^2} + \sum_{\ell=1}^3 |\nabla {S}_\ell|^2 \right) \right] \rho \,\rmd^3 x,
\label{Q-Hamiltonian-spinor}
\end{eqnarray}
where ${S}_\ell =\rho^{-1}\Psi^* \bi{{\sigma}}_\ell \Psi $ 
($\bi{{\sigma}}_\ell$ are the Pauli matrices);
explicitly, we may write
\begin{equation}
\left\{ \begin{array}{l}
{\displaystyle
S_1 := \frac{1}{\rho} \Psi^* \left( \begin{array}{cc}
0 & 1 \\
1 & 0  \end{array} \right) \Psi
% = \frac{2\sqrt{\rho_1 \rho_2}}{\rho} \cos[ (\varphi_1-\varphi_2)/\hbar] , 
= \frac{\sqrt{\rho^2-{\alpha}^2}}{\rho} \cos[2{\beta}/\hbar],
}
\\
{\displaystyle
S_2 := \frac{1}{\rho} \Psi^* \left( \begin{array}{cc}
0 & -i \\
i & 0  \end{array} \right) \Psi
% = \frac{-2\sqrt{\rho_1 \rho_2}}{\rho} \sin [(\varphi_1-\varphi_2)/\hbar] , 
=\frac{-\sqrt{\rho^2-{\alpha}^2}}{\rho} \sin[2{\beta}/\hbar],
}
\\
{\displaystyle
S_3 := \frac{1}{\rho} \Psi^* \left( \begin{array}{cc}
1 & 0 \\
0 & -1 \end{array} \right) \Psi
% = \frac{\rho_1-\rho_2}{\rho}.
=\frac{{\alpha}}{\rho} .}
\end{array} \right.
\label{spin}
\end{equation}
In the limit of $\hbar\rightarrow0$, $H_{\mathrm{q}}$ reduces to the classical fluid energy (\ref{Hamiltonian}).

The action is $\int (\Theta - {H}_{\mathrm{q}})\,\rmd t$ with
\begin{equation}
\Theta  = 
 \Re \dangle{\Psi}{i\hbar \partial_t\Psi}  
= -\int \left(\rho \partial_t\varphi + {\alpha} \partial_t{\beta} \right)\,\rmd^3 x .
\label{canonical_1-form-Clebsch}
\end{equation}
When perturbed by $\Psi$, the variational principle yields 
\begin{equation}
i\hbar\partial_t \psi_j =
 \left(- \frac{\hbar^2}{2} \nabla^2 + h \right) \psi_j
\quad (j=1,2) .
\label{Schroedinger-baroclinic}
\end{equation}
On the other hand, 
when we represent the action by the Clebsch parameters $\bxi=(\rho,\alpha,\varphi,\beta)$, 
we obtain the equivalent quantum mechanics dictated by the canonical Poisson bracket:
\begin{equation}
\dot{\bxi} = \dpoisson{\bxi}{H_{\mathrm{q}}(\bxi)}_{\mathfrak{sp}(6)}.
\label{Schroedinger-baroclinic-Clebsch}
\end{equation}
By (\ref{Madelung-2})-(\ref{Clebsch-1}), we see the equivalence of (\ref{Schroedinger-baroclinic}) and (\ref{Schroedinger-baroclinic-Clebsch}).

The \emph{Clebsch reduction} to the classical fluid bracket $\dpoisson{~}{~}_F$ is possible in the limit of $\hbar^2\rightarrow0$.
Notice that $H_{\mathrm{q}}$ of (\ref{Q-Hamiltonian-spinor}) includes terms that are not functions of the fluid variables $\bi{u}(\bxi)=(\rho,\bi{v})$;
specifically the terms including $S_\ell$.
In the limit of $\hbar^2\rightarrow0$, all remaining terms only depend on $(\rho,\bi{v})$, so that
the canonical Poisson bracket $\dpoisson{~}{~}_{\mathfrak{sp}(6)}$ is reduced to the fluid bracket $\dpoisson{~}{~}_F$.
When we retain the order $\hbar^2$ terms in $H_{\mathrm{q}}$, the ``fluid equation'' includes a quantum mechanical term;
on the right-hand side of (\ref{momentum}), we have to add
\begin{equation}
\bi{F}_q= \frac{\hbar^2}{2} \left\{ 
\nabla \left( \frac{\nabla^2\sqrt{\rho}}{\sqrt{\rho}}\right)
- \sum_{\ell=1}^3  \frac{\nabla\cdot[\rho\nabla {S}_\ell\otimes \nabla {S}_\ell]}{2\rho} \right\}.
\label{fluid_canonical-7} 
\end{equation}

\subsection{Quantum field with finite (non-exact) helicity}
\label{subsec:su(3)}
By a simple extension of the previous discussion,
we may formulate the quantization of the general fluid Poisson algebra $\dpoisson{~}{~}_F$.
The 6-component Clebsch parameters (\ref{canonical_variables}) mediate the canonical corresponding principle.
% We investigate the quantization of the finite-helicity fluid.

Here we consider 
a 3-component wave function $\Psi=(\psi_1,\psi_2,\psi_3)^{\mathrm{T}}$ which is parameterized as
\begin{equation}
\psi_j = \sqrt{\rho_j} \rme^{i\mathscr{S}_j/\hbar} \quad (j=1,2,3) .
\label{Madelug-Clebsch-1} 
\end{equation}
Extending (\ref{Clebsch-1}), we define
\begin{equation}
\left\{ \begin{array}{l}
\rho ~= \rho_1+\rho_2,+\rho_3, 
\\
{\alpha}_1 = \rho_1 - \rho_3,
\\
{\alpha}_2 = \frac{1}{2} \rho_1 - \rho_2 + \frac{1}{2} \rho_3,
\end{array} \right.
\quad
\left\{ \begin{array}{l}
\varphi ~=\frac{1}{3} (\mathscr{S}_1+\mathscr{S}_2+\mathscr{S}_3),
\\
{\beta}_1 = \frac{1}{2} (\mathscr{S}_1- \mathscr{S}_3),
\\
{\beta}_2 = \frac{1}{3} (\mathscr{S}_1- 2 \mathscr{S}_2 +  \mathscr{S}_3).
\end{array} \right.
\label{Madelug-Clebsch-2}
\end{equation}
The momentum is
\begin{equation}
\bi{v}  = \Re \frac{\Psi^* \cdot (-i\hbar \nabla\Psi) }{\rho} = 
\nabla\varphi +  \frac{{\alpha_1}}{\rho} \nabla {\beta_1}  +  \frac{{\alpha_2}}{\rho} \nabla {\beta_2},
\label{Clebsch2-2}
\end{equation}
which fits the Clebsch parameterized fluid velocity (momentum 1-form) $\bi{v}$ of (\ref{Clebsch}).
The Hamiltonian (\ref{Q-Hamiltonian-spinor}) is extended to
\begin{eqnarray}
H_{\mathrm{q}} 
% &=& \dangle{\Psi}{\mathcal{H}\Psi}
% \nonumber \\
&=&
\int \Psi^* \cdot \left[ \left(- \frac{\hbar^2}{2} \nabla^2 + \varepsilon \right) \Psi
\right]
\,\rmd^3 x 
\nonumber \\
&=& 
\int \left[ \frac{1}{2} \left| \nabla \varphi + \check{\alpha}_1\nabla{\beta}_1 +  \check{\alpha}_2\nabla{\beta}_2\right|^2 
+ \varepsilon 
\right.
\nonumber \\
& & ~~~~~~~~~~~~~~~~~~~~~~~~~~~~~ \left. +
\frac{\hbar^2}{8}  \int \left(  \frac{|\nabla\rho|^2}{\rho^2} + \sum_{\ell=1}^8 |\nabla {S}_\ell|^2 \right) \right]\rho \,\rmd^3 x,
\label{Q-Hamiltonian-su(3)}
\end{eqnarray}
where
\begin{equation}
S_\ell = \frac{1}{\rho} \Psi^* \lambda_\ell \Psi
\label{su(3)_components}
\end{equation}
with the Gell-Mann matrices
\begin{eqnarray}
& &
\lambda_1 = \left( \begin{array}{ccc}
0 & 1 & 0 \\
1 & 0 & 0 \\
0 & 0 & 0 
\end{array} \right),
\quad 
\lambda_2 = \left( \begin{array}{ccc}
0 &-i & 0 \\
i  & 0 & 0 \\
0 & 0 & 0 
\end{array} \right),
\quad 
\lambda_3 = \left( \begin{array}{ccc}
1 & 0 & 0 \\
0 &-1& 0 \\
0 & 0 & 0 
\end{array} \right),
\nonumber \\
& &
\lambda_4 = \left( \begin{array}{ccc}
0 & 0 & 1 \\
0 & 0 & 0 \\
1 & 0 & 0 
\end{array} \right),
\quad 
\lambda_5 = \left( \begin{array}{ccc}
0 & 0 &-i \\
0 & 0 & 0 \\
i & 0 & 0 
\end{array} \right),
\label{Gell-Mann}
\\
& &
\lambda_6 = \left( \begin{array}{ccc}
0 & 0 & 0 \\
0 & 0 & 1 \\
0 & 1 & 0 
\end{array} \right),
\quad 
\lambda_7 = \left( \begin{array}{ccc}
0 & 0 & 0 \\
0 & 0 &-i \\
0 & i & 0 
\end{array} \right),
\quad 
\lambda_8 = 
\frac{1}{\sqrt{3}}
\left( \begin{array}{ccc}
1 & 0 & 0 \\
0 & 1 & 0 \\
0 & 0 &-2 
\end{array} \right).
\nonumber
\end{eqnarray}
Explicitly we may write
\begin{eqnarray}
& & S_1 = \frac{2\sqrt{\rho_1\rho_2}}{\rho} \cos (\mathscr{S}_1-\mathscr{S}_2)/\hbar ,
\quad
S_2 = - \frac{2\sqrt{\rho_1\rho_2}}{\rho} \sin (\mathscr{S}_1-\mathscr{S}_2)/\hbar ,
\nonumber \\
& & S_4 = \frac{2\sqrt{\rho_1\rho_3}}{\rho} \cos (\mathscr{S}_1-\mathscr{S}_3)/\hbar ,
\quad
S_5 = - \frac{2\sqrt{\rho_1\rho_3}}{\rho} \sin (\mathscr{S}_1-\mathscr{S}_3)/\hbar ,
\label{su(3)_components-2}
\\
& & S_6 = \frac{2\sqrt{\rho_2\rho_3}}{\rho} \cos (\mathscr{S}_2-\mathscr{S}_3)/\hbar ,
\quad
S_7 = - \frac{2\sqrt{\rho_2\rho_3}}{\rho} \sin (\mathscr{S}_2-\mathscr{S}_3)/\hbar ,
\nonumber \\
& & S_3 = \frac{\rho_1 - \rho_2}{\rho},
\quad
 S_8 = \frac{1}{\sqrt{3}} \frac{\rho_1 + \rho_2 - 2 \rho_3}{\rho}.
\nonumber
\end{eqnarray}
In the limit of $\hbar^2\rightarrow0$, the canonical bracket $\dpoisson{~}{~}_{\mathfrak{sp}(6)}$ is reduced to the fluid bracket $\dpoisson{~}{~}_F$,
which has the Casimir (helicity)
\begin{eqnarray}
C &=& \int \bi{v}\cdot \nabla\times\bi{v}\,\rmd^3 x
\nonumber \\
&=& \int \left[ \frac{{\alpha}_2}{\rho}  \nabla \left(\frac{{\alpha}_1}{\rho}\right) - \frac{{\alpha}_1}{\rho}  \nabla \left(\frac{{\alpha}_2}{\rho}\right) \right]
 \cdot \nabla {\beta}_1 \times \nabla {\beta}_2 \,\rmd^3 x .
\label{helicity_Q}
\end{eqnarray}

\begin{remark}[quantization of vortex dynamics]
\label{remark:quatization}
\normalfont
Remembering the claim that the vortex dynamics is an example of classical field Nambu dynamics (Sec.\,\ref{sec:vortex_dynamics}),
we show how it can be quantized.
We first apply the Clebsch parameterization (\ref{Clebsch}) to canonicalize it on the vector bundle of $\mathfrak{sp}(6;\mathbb{R})$,
which can be related to the 3-component wave function (\ref{Madelug-Clebsch-1}).
Notice that the vortex field Clebsch parameterization (\ref{Clebsch_for_omega}) is more economical for canonicalization on the vector bundle of $\mathfrak{sp}(4;\mathbb{R})$,
however, it does not fit the Madelung representation of wave functions.
We once include $\rho$ and $\varphi$ to the phase space, and then assume that $\varphi$ and $\rho$ are constant numbers in the Hamiltonian (\ref{Q-Hamiltonian-su(3)}).
Such a subsystem is equivalent to the vortex dynamics system; see Sec.\,\ref{subsec:subsystems} (1).
We also note that the vorticity of the wave functions is different from spin.
Applying the discussion of Sec.\,\ref{subsec:su(2)}, the spin reduction of vector bundle of $\mathfrak{su}(2)$ may be viewed as an alternative realization of Nambu dynamics, which is readily quantized by spinor to obtain the system that only depends on $S_j$.
The vortex dynamics system is different; it is the curl-deformed $\mathfrak{so}(3)$ Lie-Poisson algebra (Sec.\,\ref{subsec:curl}).
\end{remark}

\begin{remark}[baroclinic effect]
\label{remark:baroclinic}
\normalfont
As mentioned in Remark\,\ref{remark:entropy}, we can consider the Hamiltonian mechanics of isentropic fluid, in which the baroclinic effect emerges. 
The entropy conservation law (Lie-dragging of the specific entropy $\sigma$), i.e., $(\partial_t + \mathcal{L}_{\bi{v}})\sigma=0$ is added to Hamilton's equation.  
One of the Clebsch parameters can play the role of $sigma$; let the scalar $\beta_1$ be the specific entropy.
In the fluid Hamiltonian (\ref{Hamiltonian}), we consider an internal energy such that $\varepsilon(\varrho^*, \beta_1)$.
Then, $\partial(\varepsilon\rho)/\partial\beta_1 = T \rho$ is added to the right-hand side of (\ref{H-3}) to produce the baroclinic effect.
The second equation of (\ref{H-2}) is as is, implying the entropy conservation law.
When such a system is quantized, an interesting effect appears in the Pauli-Schr\"odinger equation\,\cite{YoshidaMahajan2016}.
In the quantized Hamiltonian, imaginary terms (proportional to $T \rho$) are added to make it non-hermitian (in a special way so that the time-reversal symmetry is preserved),
which produces interactions between different spinor components;
the coupled Gross-Pitaevskii equations describe the baroclinic vortex dynamics in the quantum mechanical regime.
\end{remark}

\begin{remark}[EM coupling, quantum plasma]
\label{remark:EM}
\normalfont
We can include the $U(1)$ gauge field to the present formulation of classical and quantum fluid systems; the 4-momentum $(\rho,-\bi{v})$ is made ``canonical'' by adding the electromagnetic 4-potential $q (\phi, -\bi{A})$ ($q$ is the charge).
We often consider a multi-fluid system consisting of differently charged fluids (for example, electrons and ions);
then, for each fluid constituent, we apply the Clebsch parameterization.
Combined with Maxwell's equation, we can formulate a model of plasmas.
See\,\cite{Brodin,Mahajan} and papers cited there for the study of spin plasmas that may be the physical examples where the present theory applies. 
\end{remark}

\section{Corresponding principle for second quantization}
\label{sec:second_quatization}

To simplify notation, let us rewrite the Clebsch parameters as
\[
\left\{ \begin{array}{l}
q_0= \rho, 
\\
q_j = \alpha_j ,
\end{array} \right.
\quad 
\left\{ \begin{array}{l}
p_0= \varphi, 
\\
p_j = \beta_j .
\end{array} \right.
\]
We denote $\bxi(\bi{x}) = (q_0(\bi{x}),\cdots,q_{n-1}(\bi{x}), p_0(\bi{x}),\cdots,p_{n-1}(\bi{x}))^{\mathrm{T}}$.
The classical Poisson bracket $\dpoisson{~}{~}_{\mathfrak{sp}(2n)}$
will be simply denoted by $\dpoisson{~}{~}$.
We write
\[
\{ G(\bxi(\bi{x})), H(\bxi(\bi{y})) \} = \langle \partial_{\bxi} G (\bi{x}), J_c \partial_{\bxi} H (\bi{y}) \rangle ,
\]
by which $\dpoisson{G}{H} = \int_\Omega \{ G(\bxi(\bi{x})), H(\bxi(\bi{x})) \}\,\rmd^n x$;
remember (\ref{canonical_Poisson_bracket}).
The classical fields (cross-sections of the $\mathfrak{sp}(2n;\mathbb{R}))$ bundle) are dictated by the canonical commutation rule\,\cite{field_theory_braket}
% \footnote{
% The entries of the bracket $\dpoisson{~}{~}$ are functionals on the function space $C^\infty(\Omega)$.
% Here $q_j(\bi{x})$ means $\dangle{\delta(\bi{x}-\bi{x}')}{q_j(\bi{x}')}$. }
\begin{eqnarray}
& &
\dpoisson{q_j(\bi{x})}{q_k(\bi{y})} = 0,
\quad 
\dpoisson{p_j(\bi{x})}{p_k(\bi{y})} = 0,
\nonumber \\
& &\dpoisson{q_j(\bi{x})}{p_k(\bi{y})} = \delta_{jk} \delta(\bi{x}-\bi{y}), 
\quad (j,k \in \{ 0,\cdots, n-1\}, \bi{x},\bi{y}\in\Omega).
\label{classical_canonical_bracket-0}
\end{eqnarray}
By (\ref{Madelug-Clebsch-2}), or its lower-dimension representation (\ref{Clebsch-1}), (\ref{classical_canonical_bracket-0}) is equivalent to
\begin{eqnarray}
& &
\dpoisson{\rho_j(\bi{x})}{\rho_k(\bi{y})} = 0,
\quad 
\dpoisson{\mathscr{S}_j(\bi{x})}{\mathscr{S}_k(\bi{y})} = 0,
\nonumber \\
& &
\dpoisson{\rho_j(\bi{x})}{\mathscr{S}_k(\bi{y})} = \delta_{jk} \delta(\bi{x}-\bi{y}), 
\quad
(j,k \in \{ 1,\cdots, n\}, \bi{x},\bi{y}\in\Omega).
\label{classical_canonical_bracket-1}
\end{eqnarray}
About the wave function $\psi_j = \sqrt{\rho_j} \rme^{i\mathscr{S}_j/\hbar} $,
we observe
\begin{equation}
\{ \psi_j, \psi_k^* \} = 
\frac{\partial\psi_j}{\partial\rho_j} \frac{\partial\psi_k^*}{\partial\mathscr{S}_k}  \{\rho_j,\mathscr{S}_k\}
+  \frac{\partial\psi_j}{\partial\mathscr{S}_j} \frac{\partial\psi_k^*}{\partial\rho_k}  \{\mathscr{S}_j,\rho_k\} ,
\label{classical_canonical_bracket-2}
\end{equation}
and $\{ \psi_j, \psi_k \} =0$, $\{ \psi_j^*, \psi_k^* \} =0$;
hence, (\ref{classical_canonical_bracket-1}) translates as
\begin{eqnarray}
& &
\dpoisson{\psi_j(\bi{x})}{\psi_k(\bi{y})} = 0,
\quad 
\dpoisson{\psi_j^*(\bi{x})}{\psi_k^*(\bi{y})} = 0,
\nonumber \\
& &\dpoisson{\psi_j(\bi{x})}{\psi_k^*(\bi{y})} = \frac{1}{i\hbar}\delta_{jk} \delta(\bi{x}-\bi{y}) .
% \quad (j,k \in \{ 0,\cdots, n-1\}, \bi{x},\bi{y}\in\Omega).
\label{classical_canonical_bracket-3}
\end{eqnarray}
Therefore, the second-quantized Pauli-Schr\"odinger field
(acting on the Fock space of either Bosons or Fermions) is given by the corresponding principle\,\cite{Jackiw}:
\begin{eqnarray}
& & {[} \psi_j(\bi{x}), \psi_k(\bi{y}) {]}_\pm  =0, \quad
{[} \psi_j^*(\bi{x}), \psi_k^*(\bi{y}) {]}_\pm =0,
\nonumber \\
& & {[} \psi_j(\bi{x}), \psi_k^*(\bi{y}) {]}_\pm = \frac{1}{i\hbar} \delta_{jk} \delta(\bi{x}-\bi{y}) .
\label{canonical_bracket-Q}
\end{eqnarray}

% \begin{remark}[scaling]
% To separate the quantum term $\mathscr{H}_q$ from $\mathscr{H}_c$ by the order of $\hbar^2$,
% we must demand $\nabla {\beta} = O(\hbar)$.
% Then, $\bomega=\nabla\times\bi{p} =O(\hbar)$.
% This `quantal vorticity' is carried by the spin.
% In fact, we may write 
% \[
% \bomega=-\frac{\hbar}{2}\nabla S_1\times\nabla S_2 / S_3.
% \]
% \end{remark}

%%%%%%%%%%%%%%%%%%%%%%%%%%%%%%%%%%%%%%%%%%%%%%%%%%%%%%%%%%%
\section{Conclusion}
Canonical Hamiltonian mechanics describes physics theories with two ingredients: the energy (Hamiltonian) characterizing
the mechanical property of matter and symplectic geometry dictating the universal rules of kinematics in the phase space.
Usually, the variety of dynamics is ascribed to the Hamiltonian, while the geometry of phase space is maintained to be canonical.
However, deforming the geometry
of phase space (keeping the Hamiltonian simple), can be more effective for studying systems in which some topological
constraints foliate the phase space (such systems are called noncanonical). 
For example, a holonomic constraint reduces the effective space to a leaf embedded in the original canonical phase space, on which some interesting Lie algebra may dictate the kinematics. 
There are also many examples of \emph{noncanonical Hamiltonian systems} in fluid and plasma physics, where the essence of mechanics is attributed to
complex Poisson brackets.

Nambu's idea of multiple Hamiltonian systems may be viewed as such an attempt of deforming the phase space by topological constraints; the additional Hamiltonians pose constraints by the requirement of their conservation.
In that, the additional constraints may be translated as Casimirs.
However, Nambu's proposal includes some more ambitious aspect, if one assumes that the additional Hamiltonians may be freely selected,
which means that the phase space can be arbitrarily foliated by the additional Hamiltonians. 
For the system to be \emph{dynamical}, this is in fact a tall order.
As shown in Sec.\,\ref{sec:Lie-Poisson}, rather limited foliation of the phase space is allowed.
% , if we demand the fundamental relation (which translates as the Jacobi identity on the symplectic leaves).

\begin{figure}
\begin{center}
\includegraphics[width=6cm]{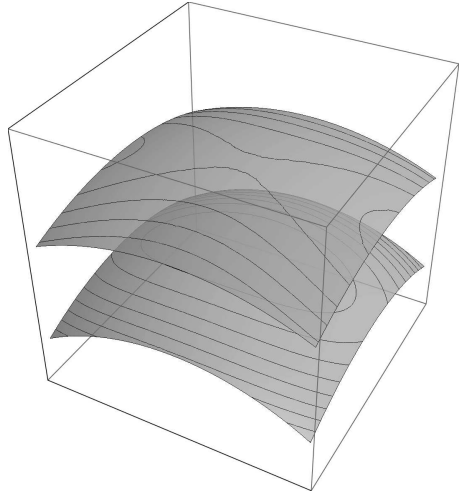}
\caption{
Foliated phase space.
The effective space on which the constrained dynamics occurs is a leaf (submanifold) embedded 
in an \emph{a priori} phase space (here $\mathbb{R}^3$).
A specific leaf is selected by the initial condition.
The effective Hamiltonian (here the Euclid norm measuring the distance from the origin)
of the constrained dynamics is determined by the shape of the leaf.
The contour lines on two different leaves show how the effective Hamiltonian changes depending on the curvature of the leaves.
}
\label{fig:foliation}
\end{center}
\end{figure}

In this work, we chose the stand point to consider that the additional Hamiltonians are Casimirs imposing specific constraints on the geometry, and discussed the way of canonicalization and quantization of such systems.
The level sets of the Casimir (which corresponds to the helicity of the classical fields) foliates the canonical phase space.
Variety of interesting structures/dynamics emerge on the leaves because the Hamiltonian, even if it is a simple convex functional, may appear to be rather complex functional when restricted on the leaf of the Casimir (Fig.\,\ref{fig:foliation}).

%---------------------------------------------------------------  FIG 1
%----------------------------------------------------------------------

As depicted in Sec.\,\ref{sec:reduction} by the simple example of $\mathfrak{so}(3)$ Lie-Poisson algebra (the example considered by Nambu), 
we can associate a Casimir with an additional conjugate variable (at least locally on the Poisson manifold), 
and embed the noncanonical Poisson manifold in the inflated canonical phase space. 
In the extended canonical system, the absence of the conjugate variable in the Hamiltonian yields a constant that is the Casimir of the original noncanonical system.
In Sec.\,\ref{sec:reduction}, we have also seen that the $\mathfrak{so}(3)$ (or $\mathfrak{su}(2)$) Lie-Poisson algebra can be derived by the angular momentum (or spin) reduction.
Conversely, such a reduced system can be canonicalized by parameterizing the physical quantities in terms of the canonical variables.
Broadly expanding the original meaning due to Clebsch, the reverse of such reduction was called Clebsch parameterization.
In fact, the advanced examples from fluid mechanics, discussed in Secs.\,\ref{sec:vortex_dynamics} and \ref{sec:general_fluid}, invoke the parameterizations close to the original idea of Clebsch.
The Casimir of the reduced Poisson algebra is the generator of the gauge group that keeps the Clebsch parameterization invariant.

The Madelung representation of complex wave functions by fluid-like real fields projects the reduction/Clebsch parameterization of Poisson algebras to Lie algebras of quantized observables, 
by which we can relate a noncanonical Hamiltonian system (Nambu dynamics) to a canonicalized Poisson algebra, and then, to a quantized system 
through the canonical corresponding principle.

% \section*{Acknowledgment}
\ack
The author thanks Professor Philip J. Morrison and Professor Swadesh M. Mahajan for their suggestions and discussions.
This review paper draws heavily on the results of collaborations with them.
The author also appreciates the private communication with Professor Toshiaki Kori on the Clebsch parameterization.
This work was partly supported by the JSPS KAKENHI under Grant No. 17H01177,
as well as by Osaka City University Advanced Mathematical Institute: MEXT Joint Usage/Research Center on Mathematics
and Theoretical Physics JPMXP0619217849.

% \let\doi\relax

%without this code before the command "\begin{thebibliography}{}" , an error will be %flagged. When the bibliography is provided as separate .bib file, then this code %should be placed above the commands "\bibliographystyle{}" and "\bibliography{}" %inside the main TeX file. 

\begin{thebibliography}{9}
\bibitem{covariant}
In Sec.\,\ref{sec:introduction}, we did not distinguish the contravariant and covariant variables.
In the following discussions, the duality of both vectors will be important;
$\mathfrak{g}$ is the space of contravariant vectors (operators), while $\mathfrak{g}^*$ is the space of covariant vectors (observables).
\bibitem{semi-simple}
For a semi-simple Lie algebra (like $\mathfrak{so}(3)$), we can formally evaluate $[\bi{x},\bi{y}]^* = [\bi{x},\bi{y}]$.
This means that we may identify $\mathfrak{g}=\mathfrak{g}^*$ with an appropriate basis of $\mathfrak{g}^*$,
and that the structure constants are \emph{fully antisymmetric}; for example, see \cite{YM2020}.
We note that the general dual bracket $[~,~]^*$ is not necessarily a Lie bracket 
(even $[\bxi,\bxi]^* =0$ may not hold).
\bibitem{Heisenberg}
As seen in the general construction of the Lie-Poisson bracket (Sec.\,\ref{subsec:general_Lie-Poisson}),
the Poisson operator $J$ is a linear function of the phase space coordinates.
In (\ref{Heisenberg-Poisson}),  $J$ depends on $r$, but $r$ is a Casimir, so it is constant on each symplectic leaf.
\bibitem{Hamilton_eq}
Comparing $\dot{G}=\dangle{\partial_{\bomega} G }{\dot{\bomega}}$ and
$\dot{G}=\dpoisson{G}{H}_{\bomega} = \dangle{\partial_{\bomega} G}{J_{\bomega} \partial_{\bomega} H}$
($\forall G(\bomega)$), 
we obtain (\ref{vortex_eq-Hamiltonian}) as an evolution equation in the topology of $L^2(\Omega)$.
\bibitem{compressible}
This is not the case for compressible fluids (see Sec.\,\ref{sec:general_fluid}).
\bibitem{Clebsch_forms}
Here the components of $\bxi \in X$ are mixture of 0-forms and 3-forms (for the convenience of the latter formulation).  So the duality is defined in a different way in comparison with the definition in Remark\,\ref{remark:duality}.
\bibitem{generator}
The temporal derivative $\dot{\mathcal{G}}$ (or $\dot{G}$) is evaluated at $t=0$.
\bibitem{nonlinearity}
Notice that $H_{\mathrm{q}}$ of (\ref{fluid_energy-q}) is not $\dangle{\psi}{\mathcal{H}\psi}$, reflecting the nonlinearity due to $\varepsilon(|\psi|^2)$.
\bibitem{field_theory_braket}
The entries of the bracket $\dpoisson{~}{~}$ are functionals on the function space $C^\infty(\Omega)$.
Here $q_j(\bi{x})$ means $\dangle{\delta(\bi{x}-\bi{x}')}{q_j(\bi{x}')}$.
\end{thebibliography}

% \appendix

% \section{Appendix head}

\end{document}